\documentclass[a4paper,11pt]{article}
\usepackage[left=20mm,right=20mm,top=20mm,bottom=20mm]{geometry}
\usepackage{amsmath}
\usepackage{bm}
\usepackage{amsfonts}
\usepackage{amssymb}
\usepackage{amsbsy}
\usepackage{color,graphics}
\usepackage{setspace}
\usepackage{epsfig}
\usepackage{cite}
\usepackage{graphicx}
\usepackage{latexsym,multicol}
\usepackage{appendix}
\newtheorem{theorem}{Theorem}[section]

\graphicspath{{./SET-figs/}}
\begin{document}
\title{Likely chirality of stochastic anisotropic hyperelastic tubes}
\author{L. Angela Mihai\footnote{School of Mathematics, Cardiff University, Senghennydd Road, Cardiff, CF24 4AG, UK, Email: \texttt{MihaiLA@cardiff.ac.uk}}
	\qquad Thomas E. Woolley\footnote{School of Mathematics, Cardiff University, Senghennydd Road, Cardiff, CF24 4AG, UK, Email: \texttt{WoolleyT1@cardiff.ac.uk}}
	\qquad Alain Goriely\footnote{Mathematical Institute, University of Oxford, Woodstock Road, Oxford, OX2 6GG, UK, Email: \texttt{goriely@maths.ox.ac.uk}}
}	
\maketitle

\begin{abstract}
When an elastic tube reinforced with helical fibres is inflated, its ends rotate. In large deformations, the amount and chirality of rotation is highly non-trivial, as it depends on the choice of strain-energy density and the arrangements of the fibres. For anisotropic hyperelastic tubes where the material parameters are single-valued constants, the problem has been satisfactorily addressed. However, in many systems, the material parameters are not precisely known, and it is therefore more appropriate to treat them as random variables. The problem is then to understand chirality in a probabilistic framework. Here, we develop a procedure for examining the elastic responses of a hyperelastic cylindrical tube of stochastic anisotropic material, where the material parameters are spatially-independent random variables defined by probability density functions. The tube is subjected to uniform dead loading consisting of internal pressure, axial tension and torque. Assuming that the tube wall is thin and that the resulting deformation is the combined inflation, extension and torsion from the reference circular cylindrical configuration to a deformed circular cylindrical state, we derive the probabilities of radial expansion or contraction, and of right-handed or left-handed torsion. We refer to these stochastic behaviours as `likely inflation' and `likely chirality', respectively.\\

\end{abstract}

\section{Introduction}

Biological, medical, and engineering applications of the extension-torsion-inflation couplings, in pressurised circular tubes with helical aeolotropy, range from plant and animal tissues \cite{Bertinetti:2013:BFF,Kroeger:2018:KKSS}, to cardiac and vascular systems \cite{Gilchrist:2017:GMPS,Quarteroni:2017:QLRRB}, to soft actuators \cite{Connolly:2015:CPWB,Connolly:2017:CWB} and dielectric elastomers \cite{He:2018.HLD}. The mechanical effects of the stiffness and orientation of the fibres winding helically around the tube's axis were analysed in detail, within the framework of finite elasticity, in \cite{Goriely:2013:GT}. There it was shown that: (i) the tube can lengthen or increase in girth depending on the fibre orientation and the relative stiffness of the matrix and fibres; (ii) when the fibres are equal in stiffness and orientation, the tube may exhibit no rotation; (iii) a tube with a single pre-compressed right-handed fibre may rotate either clockwise or anti-clockwise depending on the material parameters, showing that transfer of chirality under external loads from the tube-wall micro-structure to the tubular macro-structure relies both on geometry and material properties.

In addition, for many solid materials, a crucial part in assessing their physical properties is to quantify the uncertainties in their mechanical responses, which cannot be ignored \cite{Hughes:2010:HH,Oden:2018,Ostroja:2007}. In finite elasticity, the use of the variability in observational data and information about uncertainties \cite{Ghanem:2017:GHOR,Sullivan:2015} were proposed in \cite{Staber:2015:SG,Staber:2016:SG,Staber:2017:SG} and \cite{Mihai:2018:MWG}, where stochastic isotropic hyperelastic models are described by a strain-energy function, with the parameters defined as random variables characterised by probability distributions, while anisotropic stochastic models, where the parameters are spatially-dependent random fields, were discussed in \cite{Staber:2018:SG,Staber:2019:SGSMI}. These are phenomenological models, based on the notion of entropy (or uncertainty) \cite{Shannon:1948} (see also \cite{Soni:2017:SG}) and the maximum entropy principle for a discrete probability distribution \cite{Jaynes:1957a,Jaynes:1957b,Jaynes:2003}, which can also be embedded in Bayesian methodologies \cite{Bayes:1763} (see also \cite{McGrayne:2012}) for model selection and updates \cite{Mihai:2018:MWG,Oden:2018,Robert:2007}.

For these stochastic hyperelastic models, the mathematical question arises: \emph{what is the influence of probabilistic model parameters on the predicted elastic responses?} This question has begun to be addressed analytically in \cite{Mihai:2018a:MDWG,Mihai:2018b:MDWG,Mihai:2019a:MWG}, with the special cases of stochastic hyperelastic bodies with simple geometries for which the finite deformation is known. Examples include the cavitation of a sphere under uniform tensile dead load \cite{Mihai:2018a:MDWG}, the inflation of pressurised spherical and cylindrical shells \cite{Mihai:2018b:MDWG}, and the classical problem of the Rivlin cube \cite{Mihai:2019a:MWG}. For these fundamental problems, to which the elastic solution is known explicitly, the sensitive dependence on parameter probabilities was demonstrated by showing that, in contrast to the deterministic problem, where a single critical value strictly separates the cases where instability can or cannot occur, for the stochastic problem, there is a probabilistic interval where the two cases compete, in the sense that both have a quantifiable chance to be found. Such problems can offer significant insight into how the stochastic framework builds on the finite elasticity theory, and revisiting them from the stochastic perspective can offer also opportunities for gaining new insights into the elastic solutions. More complex, but mathematically tractable problems can then be treated in a similar manner \cite{Mihai:2019:MDWG}. To study the effect of probabilistic parameters in the case of more complex geometries and loading conditions, suitable numerical approaches built on rigorous analysis have started to be developed in \cite{Staber:2018:SG,Staber:2019:SGSMI}.
	
In this paper, we examine the elastic responses under dead loading of a hyperelastic cylindrical tube of stochastic anisotropic material, where the model parameters are random variables, which are constant in space, and hence are particular cases of those treated in \cite{Staber:2018:SG,Staber:2019:SGSMI}. One can regard this stochastic tube as an ensemble of tubes with the same geometry, where each individual tube is made from a homogeneous anisotropic incompressible hyperelastic material, with the elastic parameters drawn from known probability distributions. Then, for each individual hyperelastic tube in the ensemble, the finite elasticity theory applies. The dead loading consists of simultaneous internal pressure, axial tension and torque, and the resulting universal deformation is a combined inflation, extension and torsion from the reference circular cylindrical configuration to a deformed, also circular cylindrical, state. For the anisotropic material, two non-orthogonal preferred directions are assumed, corresponding to the (mean) directions of two families of aligned fibres embedded in an isotropic matrix material. In the deterministic elastic case, the tube may undergo `inversion' in the deformation, such that the radius first decreases and then increases, or `perversion' whereby the torsion chirality changes from right-handed to left-handed \cite{gota98b,mcgo02}. The possible existence of such responses depends on the material constitutive model. For the stochastic problem, we derive the probability distribution of the deformations, and find that, due to the probabilistic nature of the material parameters, the different states always compete. In particular, at a critical load, the radius may decrease or increase with a given probability, and we refer to this phenomenon as `likely inflation', and similarly, right-handed or left-handed torsion may occur with a given probability, and we refer to that  as `likely chirality'. 

The outline of this paper is as follows: in Section~\ref{sec:eit}, the kinematics of finite simple torsion superposed on uniform stretch for hyperelastic circular cylindrical tubes is briefly reviewed; Section~\ref{sec:models} provides a summary of the stochastic elasticity modelling framework; in Section~\ref{sec:tube}, the probabilistic solution for stochastic hyperelastic tubes is obtained, while some additional technical details are contained in Appendix~\ref{sec:append}; concluding remarks are presented in Section~\ref{sec:conclude}.

\section{Stretch and torsion of a circular cylindrical tube}\label{sec:eit}

\begin{figure}[htbp]
	\begin{center}
		\includegraphics[width=0.8\textwidth]{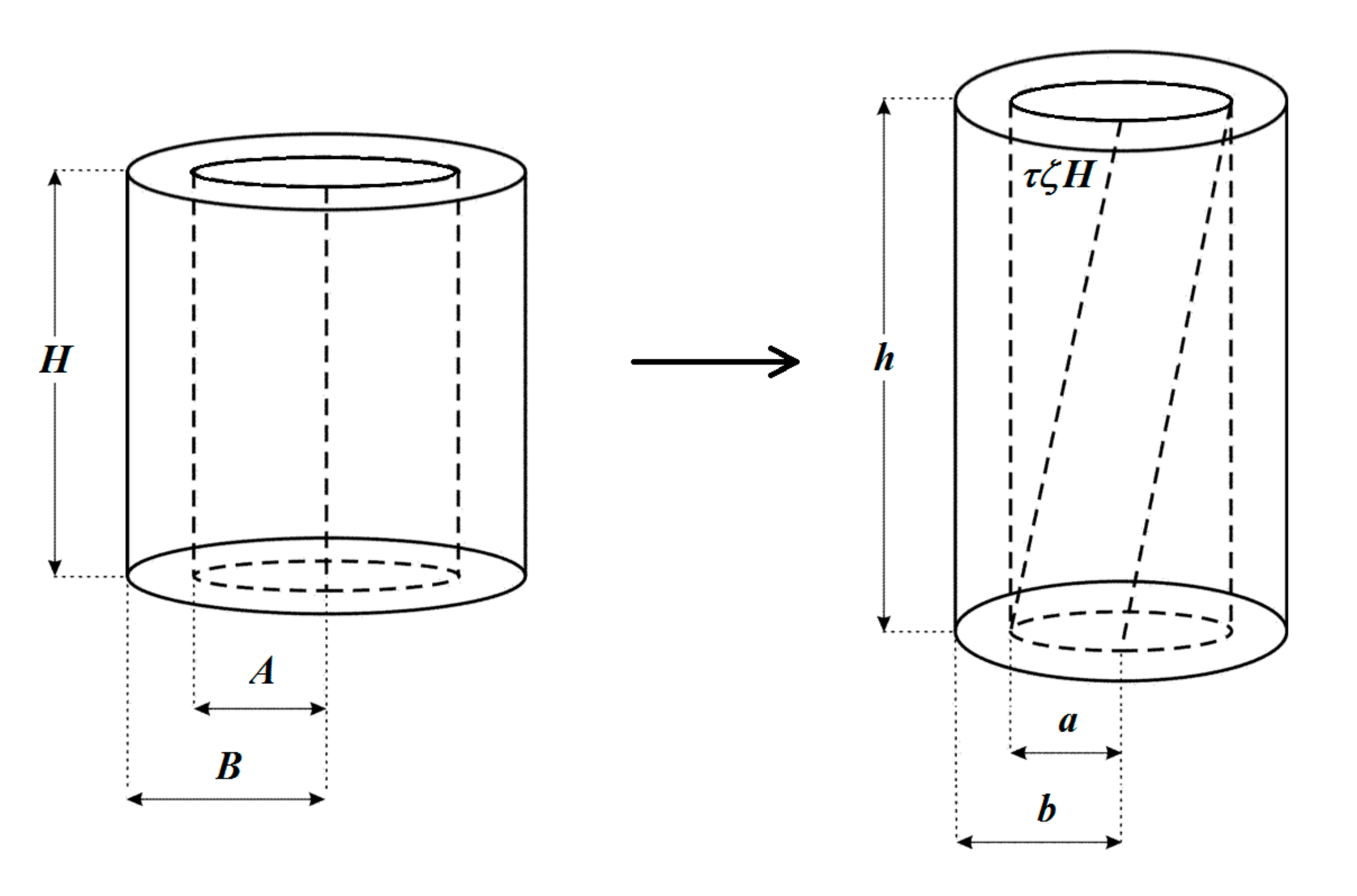}
		\caption{Schematic of circular cylindrical tube (left) deformed by combined stretch and torsion (right).}\label{fig:tube-cell}
	\end{center}
\end{figure}

We consider a circular cylindrical tube, occupying the reference domain $(R,\Theta,Z)\in[A,B]\times[-\pi,\pi)\times[0,H]$, where $A$, $B$ and $H$ are positive constants, subject to the following combined deformation consisting of simple torsion superposed on on axial stretch  \cite[pp.~184-186]{TruesdellNoll:2004}, 
\begin{equation}\label{eq:st:tube}
r=\sqrt{a^2+\frac{R^2-A^2}{\zeta}},\qquad \theta=\Theta+\tau\zeta Z,\qquad z=\zeta Z,
\end{equation}
where $(r,\theta,z)\in[a,b]\times[-\pi,\pi)\times[0,h]$ are the cylindrical polar coordinates in the deformed configuration, $a$, $\tau$ and $\zeta$ are given positive constants, $b=\sqrt{a^2+(B^2-A^2)/\zeta}$, and $h=\zeta H$ (see Figure~\ref{fig:tube-cell}).

Through the deformation \eqref{eq:st:tube}, the circular plane section at $Z=0$ remains fixed and each circular plane section normal to the central axis remains plane and rotates by an angle $\tau\zeta Z$. Denoting
\begin{equation}\label{eq:lambda}
\lambda=\frac{r}{R}=\frac{1}{R}\sqrt{a^2+\frac{R^2-A^2}{\zeta}},
\end{equation}
the deformation gradient in terms of the current cylindrical polar coordinates $(r,\theta,z)$ is equal to
\begin{equation}\label{eq:F}
\textbf{F}=\left[
\begin{array}{ccc}
\partial r/\partial R & 0 & 0\\
0 & (r/R)\partial\theta/\partial\Theta & r\partial\theta/\partial Z\\
0 & 0 & \partial z/\partial Z
\end{array}
\right]
=\left[
\begin{array}{ccc}
1/\left(\lambda\zeta\right) & 0 & 0\\
0 & \lambda & \tau\zeta r\\
0 & 0 & \zeta
\end{array}
\right],
\end{equation}
and the left and right Cauchy-Green tensors are, respectively,
\begin{eqnarray}
&&\textbf{B}=\textbf{F}\textbf{F}^T
=\left[
\begin{array}{ccc}
1/\left(\lambda^2\zeta^2\right) & 0 & 0\\
0 & \lambda^2+\tau^2\zeta^2r^2 & \tau\zeta^2r\\
0 & \tau\zeta^2r & \zeta^2
\end{array}
\right],\label{eq:B}\\
&&\textbf{C}=\textbf{F}^T\textbf{F}
=\left[
\begin{array}{ccc}
1/\left(\lambda^2\zeta^2\right) & 0 & 0\\
0 & \lambda^2 & \tau\zeta r\lambda\\
0 & \tau\zeta r\lambda & \zeta^2\left(\tau^2r^2+1\right)
\end{array}
\right].\label{eq:C}
\end{eqnarray}
Denoting $\lambda_{a}=a/A$, by \eqref{eq:lambda}, we can write
\begin{equation}\label{eq:lambda:b}
\lambda_{b}=\frac{b}{B}=\frac{1}{B}\sqrt{a^2+\frac{B^2-A^2}{\zeta}}
\end{equation}
and
\begin{equation}\label{eq:rlambda}
r=A\lambda\sqrt{1-\lambda_{a}^2\zeta}\left(1-\lambda^2\zeta\right).
\end{equation}
Next, assuming that the components of the Cauchy stress, $\textbf{T}$, are independent of $\theta$ and $z$, the equilibrium equations are reduced to \cite[p.~185]{TruesdellNoll:2004},
\begin{equation}\label{eq:equilib}
\begin{split}
\frac{\partial T_{rr}}{\partial r}+\frac{T_{rr}-T_{\theta\theta}}{r}&=0,\\
\frac{\partial T_{r\theta}}{\partial r}+2\frac{T_{r\theta}}{r}&=0,\\
\frac{\partial T_{rz}}{\partial r}+\frac{T_{rz}}{r}&=0.
\end{split}
\end{equation}
Then, by further assuming that $T_{r\theta}=0$ and $T_{rz}=0$, it follows that
\begin{equation}\label{eq:equilib:r}
\frac{\text{d}T_{rr}}{\text{d} r}=\frac{T_{\theta\theta}-T_{rr}}{r},
\end{equation}
hence,
\begin{equation}\label{eq:Ttheta}
T_{\theta\theta}=\frac{\text{d}\left(rT_{rr}\right)}{\text{d} r}.
\end{equation}
In this case, when a uniform internal pressure is applied, the outer surface of the cylinder may be rendered free of traction, i.e., $T_{rr}=0$ on $r=B$. Thus, for an internally pressurised tube which is free on the outer surface,
\begin{equation}\label{eq:load:r}
T_{rr}|_{r=a}=-P,\qquad  T_{rr}|_{r=b}=0,
\end{equation}
where $P>0$ is constant. Integration of equation  \eqref{eq:equilib:r}, with respect to $r$, followed by substitution in the first condition of \eqref{eq:load:r}, then gives
\begin{equation}\label{eq:load:P}
P=\int_{a}^{b}\frac{T_{\theta\theta}-T_{rr}}{r}dr.
\end{equation}
For the deformation \eqref{eq:st:tube}, by \eqref{eq:Ttheta}, the resultant normal force acting upon the plane $a\leq r\leq b$, $z$-constant, $|\theta|\leq \pi$ is calculated as follows \cite[pp.~185]{TruesdellNoll:2004},
\begin{equation}\label{eq:load:N}
\begin{split}
N&=2\pi\int_{a}^{b}T_{zz}rdr,\\
&=2\pi\int_{a}^{b}\left(T_{zz}-T_{rr}\right)rdr+2\pi\int_{a}^{b}T_{rr}rdr,\\
&=2\pi\int_{a}^{b}\left(T_{zz}-T_{rr}\right)rdr+\pi\int_{a}^{b}T_{rr}r^2dr+\pi\int_{a}^{b}\left(T_{rr}-T_{\theta\theta}\right)rdr,\\
&=\pi Pa^2+\pi\int_{a}^{b}\left(2T_{zz}-T_{rr}-T_{\theta\theta}\right)rdr.
\end{split}
\end{equation}
Equivalently,
\begin{equation}\label{eq:load:NF}
N=\pi Pa^2+F,
\end{equation}
where
\begin{equation}\label{eq:load:F}
F=\pi\int_{a}^{b}\left(2T_{zz}-T_{rr}-T_{\theta\theta}\right)rdr.
\end{equation}
Similarly, the resulting twisting moment is equal to \cite[p.~190]{TruesdellNoll:2004},
\begin{equation}\label{eq:load:T}
T=2\pi\int_{a}^{b}T_{\theta z}r^2dr.
\end{equation}
After applying the change of variable \eqref{eq:rlambda}, the expressions of $P$, $F$, and $T$, given by \eqref{eq:load:P}, \eqref{eq:load:F}, and \eqref{eq:load:T}, respectively, take the following equivalent forms,
\begin{eqnarray}
P&=&\int_{\lambda_{a}}^{\lambda_{b}}\frac{T_{\theta\theta}-T_{rr}}{\lambda\left(1-\lambda^2\zeta\right)}d\lambda,\label{eq:load:P:lambda}\\
F&=&\pi A^2\int_{\lambda_{a}}^{\lambda_{b}}\lambda\left(2T_{zz}-T_{rr}-T_{\theta\theta}\right)\frac{1-\lambda_{a}^2\zeta}{\left(1-\lambda^2\zeta\right)^2}d\lambda,\label{eq:load:F:lambda}\\
T&=&2\pi A^3\int_{\lambda_{a}}^{\lambda_{b}}\lambda^2T_{\theta z}\frac{\left(1-\lambda_{a}^2\zeta\right)^{3/2}}{\left(1-\lambda^2\zeta\right)^{5/2}}d\lambda.\label{eq:load:T:lambda}
\end{eqnarray}

\section{Stochastic anisotropic hyperelastic material}\label{sec:models}

In this section, we combine finite elasticity  \cite{goriely17,Ogden:1997,TruesdellNoll:2004} and probability theory \cite{Grimmett:2001:GS,Jaynes:2003} to analyse a tube made of a stochastic homogeneous hyperelastic material for which the model parameters are random variables characterised by probability density functions. For each model parameter, the partial information provided by the  mean value and variance is commonly used \cite{Caylak:2018:etal,Hughes:2010:HH,McCoy:1973,Norenberg:2015:NM}. Specifically, we assume that the tube is made of an aeolotropic material with two preferred directions with respect to the reference configuration, induced by two families of aligned extensible fibres embedded in an isotropic matrix material. The two preferred directions are taken as \cite{Goriely:2013:GT}, \cite[pp.~328-336]{goriely17}
\begin{equation}\label{eq:M12}
\textbf{M}_{1}
=\left[
\begin{array}{c}
M_{1r}\\
M_{1\theta}\\
M_{1z}
\end{array}
\right]
=\left[
\begin{array}{c}
0\\
\cos\Phi\\
\sin\Phi
\end{array}
\right]
\qquad\mbox{and}\qquad
\textbf{M}_{2}
=\left[
\begin{array}{c}
M_{2r}\\
M_{2\theta}\\
M_{2z}
\end{array}
\right]
=\left[
\begin{array}{c}
0\\
-\cos\Psi\\
\sin\Psi
\end{array}
\right],
\end{equation}
where $\Phi,\Psi\in[0,\pi/2]$. Under the deformation \eqref{eq:st:tube}, the stretch ratios of the fibres, $\lambda_{4}$ and $\lambda_{6}$, respectively, are given by
\begin{equation}\label{eq:I46}
I_{4}=\lambda_{4}^2=\left(\textbf{C}\textbf{M}_{1}\right)\cdot\textbf{M}_{1},\qquad
I_{6}=\lambda_{6}^2=\left(\textbf{C}\textbf{M}_{2}\right)\cdot\textbf{M}_{2},
\end{equation}
where $\textbf{C}$ is the right Cauchy-Green tensor defined by \eqref{eq:C}.

For the corresponding homogeneous hyperelastic model, the strain-energy function requires seven independent invariants \cite{Aguiar:2018a:ALdR,Aguiar:2018b:ALdR}, including the principal invariants, $I_{1}$, $I_{2}$, $I_{3}$, of the Cauchy-Green tensors $\textbf{B}$ and $\textbf{C}$ \cite{Spencer:1971}, the pseudo-invariants $I_{4}$ and $I_{6}$ given by \eqref{eq:I46}, and two other pseudo-invariants defined as follows,
\begin{equation}\label{eq:I57}
I_{5}=\left(\textbf{C}^2\textbf{M}_{1}\right)\cdot\textbf{M}_{1},\qquad
I_{7}=\left(\textbf{C}^2\textbf{M}_{2}\right)\cdot\textbf{M}_{2}.
\end{equation}
For a review of a wide range of isotropic and anisotropic hyperelastic models, we refer, for example, to \cite{Chagnon:2014:CRF}.

Here, we restrict our attention to the simple case of a stochastic incompressible anisotropic hyperelastic material with two families of extensible fibres embedded in a neo-Hookean material. Hence, we define the following strain-energy function,
\begin{equation}\label{eq:W:stoch}
\mathcal{W}(I_{1},I_{4},I_{6})=\frac{\mu}{2}\left(I_{1}-3\right)+\frac{\mu_{4}}{4}\left(I_{4}-1\right)^2+\frac{\mu_{6}}{4}\left(I_{6}-1\right)^2,
\end{equation}
where $\mu$, $\mu_{4}$ and $\mu_{6}$ are positive random variables, which we assume to be stochastically independent \cite{Staber:2018:SG,Staber:2019:SGSMI}. This assumption enables us to extend directly the analytical calculations from the deterministic case treated in \cite{Goriely:2013:GT}, where $\mu$, $\mu_{4}$ and $\mu_{6}$ were single-valued constants, to the stochastic problem, where these parameters are characterised by probability distributions. From the continuum mechanics perspective, in general, anisotropic materials have distinct material properties in different directions. For our model example, the shear moduli are $\mu$ in two directions and a linear combination of $\mu$, $\mu_{4}$ and $\mu_{6}$ in the third direction. Therefore, assuming that $\mu$, $\mu_{4}$, and $\mu_{6}$ are independent random variables allows for the shear moduli in different directions to be derived as shown in Appendix~\ref{sec:append}. The shear moduli of other anisotropic hyperelastic materials can be treated analogously. However, the physical behaviour of stochastic anisotropic materials deserves further attention, and we hope that our theoretical analysis may serve as a motivation for future experimental work.

For the stochastic model described by \eqref{eq:W:stoch}, the following mathematical constraints guarantee that the random shear modulus of the matrix material under infinitesimal deformations, $\mu$, and its inverse, $1/\mu$, are second-order random variables, i.e., they have finite mean value and finite variance \cite{Staber:2015:SG,Staber:2016:SG,Staber:2017:SG,Mihai:2018:MWG,Mihai:2018a:MDWG,Mihai:2018b:MDWG,Mihai:2019a:MWG},
\begin{eqnarray}\label{eq:Emu1}\begin{cases}
E\left[\mu\right]=\underline{\mu}>0,&\\
E\left[\log\ \mu\right]=\nu,& \mbox{such that $|\nu|<+\infty$}.\label{eq:Emu2}\end{cases}
\end{eqnarray}
Then, by the maximum entropy principle, the random shear modulus, $\mu>0$, with mean value $\underline{\mu}$ and variance $\text{Var}[\mu]$, follows a Gamma probability distribution \cite{Soize:2000,Soize:2001}, with shape parameter $\rho_{1}>0$ and scale parameter $\rho_{2}>0$ defined, respectively, by
\begin{equation}\label{eq:rho12}
\rho_{1}=\frac{\underline{\mu}^2}{\text{Var}[\mu]},\qquad
\rho_{2}=\frac{\text{Var}[\mu]}{\underline{\mu}}.
\end{equation}
The corresponding probability density function takes the form \cite{Abramowitz:1964,Johnson:1994:JKB}
\begin{equation}\label{eq:mu:gamma}
g(\mu;\rho_{1},\rho_{2})=\frac{\mu^{\rho_{1}-1}e^{-\mu/\rho_{2}}}{\rho_{2}^{\rho_{1}}\Gamma(\rho_{1})},\qquad\mbox{for}\ \mu>0\ \mbox{and}\ \rho_{1}, \rho_{2}>0,
\end{equation}
where $\Gamma:\mathbb{R}^{*}_{+}\to\mathbb{R}$ is the complete Gamma function
\begin{equation}\label{eq:gamma}
\Gamma(z)=\int_{0}^{+\infty}t^{z-1}e^{-t}\text dt.
\end{equation}
Similarly, by setting the mathematical expectations \cite{Staber:2018:SG}
\begin{eqnarray}\label{eq:Emu41}\begin{cases}
E\left[\mu_{4}\right]=\underline{\mu}_{4}>0,&\\
E\left[\log\ \mu_{4}\right]=\nu_{4},& \mbox{such that $|\nu_{4}|<+\infty$},\label{eq:Emu42}\end{cases}
\end{eqnarray}
and
\begin{eqnarray}\label{eq:Emu61}\begin{cases}
E\left[\mu_{6}\right]=\underline{\mu}_{6}>0,&\\
E\left[\log\ \mu_6\right]=\nu_6,& \mbox{such that $|\nu_6|<+\infty$},\label{eq:Emu62}\end{cases}
\end{eqnarray}
the random parameters $\mu_{4}>0$ and $\mu_{6}>0$, with mean values $\underline{\mu}_{4}$ and $\underline{\mu}_{6}$, and variance $\text{Var}[\mu_{4}]$ and $\text{Var}[\mu_6]$, respectively, follow the Gamma probability distributions with shape and scale parameters defined, respectively, by
\begin{equation}\label{eq:rho12:mu4}
\rho_{1}^{(4)}=\frac{\underline{\mu}_{4}^2}{\text{Var}[\mu_{4}]},\qquad
\rho_{2}^{(4)}=\frac{\text{Var}[\mu_{4}]}{\underline{\mu}_{4}},
\end{equation}
and
\begin{equation}\label{eq:rho12:mu6}
\rho_{1}^{(6)}=\frac{\underline{\mu}_{6}^2}{\text{Var}[\mu_{6}]},\qquad
\rho_{2}^{(6)}=\frac{\text{Var}[\mu_{6}]}{\underline{\mu}_{6}}.
\end{equation}
We denote by $g_{4}\left(u;\rho^{(4)}_{1},\rho^{(4)}_{2}\right)$ and $g_{6}\left(u;\rho^{(4)}_{1},\rho^{(4)}_{2}\right)$ the corresponding probability distributions.

Throughout this paper, computer simulations were  created by fixing the parameters given in each figure caption, and repeatedly drawing random samples from the underlying distribution. Our simulations were run in Matlab 2018a, where we made specific use of inbuilt functions for random number generation. Namely, we used ``gamrnd'' to generate the Gamma distributed random variables and ``gamcdf''  to generate the Gamma cumulative distribution function.

\section{Inflation and torsion of a stochastic anisotropic tube}\label{sec:tube}

For a cylindrical tube of stochastic anisotropic hyperelastic material, with the strain-energy density function described by \eqref{eq:W:stoch}, and subject to the deformation (\ref{eq:st:tube}), the Cauchy stress tensor takes the form
\begin{equation}\label{eq:stress}
\textbf{T}=-p\textbf{I}+\beta_{1}\textbf{B}+\beta_{4}\textbf{F}\textbf{M}_{1}\otimes\textbf{F}\textbf{M}_{1}+\beta_{6}\textbf{F}\textbf{M}_{2}\otimes\textbf{F}\textbf{M}_{2},
\end{equation}
where $\textbf{B}$ is the left Cauchy-Green tensor described by \eqref{eq:B}, $\beta_{i}=2\partial\mathcal{W}/\partial I_{i}$, $i=1,4,6$, are the material response coefficients, and $p$ is the Lagrangian multiplier for the incompressibility constraint, $\det\textbf{F}=1$. The preferred directions given by \eqref{eq:M12} are deformed into the following directions, respectively,
\begin{equation}\label{eq:M12:st}
\textbf{m}_{1}=\textbf{F}\textbf{M}_{1}
=\left[
\begin{array}{c}
0\\
\lambda\cos\Phi+\tau\zeta r\sin\Phi\\
\zeta\sin\Phi
\end{array}
\right],\qquad
\textbf{m}_{2}=\textbf{F}\textbf{M}_{2}
=\left[
\begin{array}{c}
0\\
-\lambda\cos\Psi+\tau\zeta r\sin\Psi\\
\zeta\sin\Psi
\end{array}
\right].
\end{equation}
Thus, the non-zero components of the stress tensor given by \eqref{eq:stress} take the form,
\begin{equation}\label{eq:stress:nonzero}
\begin{split}
T_{rr}&=-p+\frac{\beta_{1}}{\lambda^2\zeta^2},\\
T_{\theta\theta}&=-p+\beta_{1}\left(\lambda^2+\tau^2\zeta^2r^2\right)+\beta_{4}\left(\lambda\cos\Phi+\tau\zeta r\sin\Phi\right)^2+\beta_{6}\left(\lambda\cos\Psi-\tau\zeta r\sin\Psi\right)^2,\\
T_{\theta z}&=\beta_{1}\tau\zeta^2r+\beta_{4}\zeta\sin\Phi\left(\lambda\cos\Phi+\tau\zeta r\sin\Phi\right)-\beta_{6}\zeta\sin\Psi\left(\lambda\cos\Psi-\tau\zeta r\sin\Psi\right),\\
T_{zz}&=-p+\beta_{1}\zeta^2+\beta_{4}\zeta^2\sin^2\Phi+\beta_{6}\zeta^2\sin^2\Psi.
\end{split}
\end{equation}

\begin{figure}[htbp]
	\begin{center}
		\includegraphics[width=0.35\textwidth]{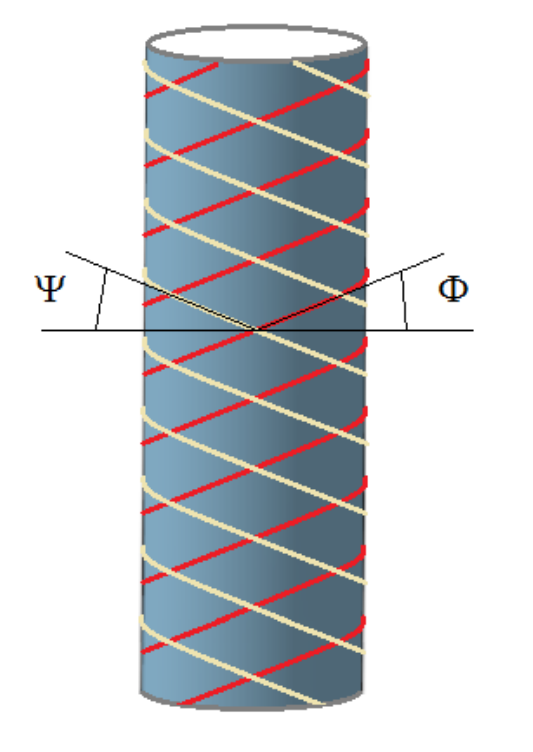}
		\caption{Schematic of cylindrical shell of anisotropic material, showing the orientation of the preferred directions induced by two families of aligned fibres tangential to the cylindrical surface.}\label{fig:anisotube}
	\end{center}
\end{figure}

Assuming that the tube wall is thin (see Figure~\ref{fig:anisotube}), we set $A=1$ and $B=1+\epsilon$, and represent $P$, $F$, and $T$, given by \eqref{eq:load:P:lambda}, \eqref{eq:load:F:lambda}, and \eqref{eq:load:T:lambda}, respectively, as the following series expansions \cite{Goriely:2013:GT},
\begin{eqnarray}
P&=&P^{(0)}+P^{(1)}\epsilon+P^{(2)}\epsilon^2+\cdots,\label{eq:load:P:app} \\
F&=&F^{(0)}+F^{(1)}\epsilon+F^{(2)}\epsilon^2+\cdots,\label{eq:load:F:app} \\
T&=&T^{(0)}+T^{(1)}\epsilon+T^{(2)}\epsilon^2+\cdots,\label{eq:load:T:app}
\end{eqnarray}
where we assume $P^{(0)}=F^{(0)}=T^{(0)}=0$. We then truncate the series given by \eqref{eq:load:P:app}, \eqref{eq:load:F:app}, and \eqref{eq:load:T:app}, respectively, to first order in $\epsilon$, as follows,
\begin{eqnarray}
P&=&\frac{1}{\zeta}\left[\mu\left(\zeta^2\tau^2+1-\frac{1}{\lambda^4\zeta^2}\right)+\mu_{4}J_{4}\left(\cos\Phi+\tau\zeta\sin\Phi\right)^2+\mu_{6}J_{6}\left(\cos\Psi-\tau\zeta\sin\Psi\right)^2\right],\label{eq:load:P:lin}\\
F&=&-\frac{\pi}{\zeta}\left[\mu\left(\lambda^2\zeta^2\tau^2-2\zeta^2+\lambda^2+\frac{1}{\lambda^2\zeta^2}\right)\right.\\
&&\left.+\mu_{4}J_{4}\left(J_{4}+1-3\zeta\sin^2\Phi\right)+\mu_{6}J_{6}\left(J_{6}+1-3\zeta\sin^2\Psi\right)\right],\label{eq:load:F:lin}\\
T&=&2\pi\lambda\left[\mu\lambda\zeta\tau+\mu_{4}J_{4}\lambda\sin\Phi\left(\cos\Phi+\zeta\tau\sin\Phi\right)+\mu_{6}J_{6}\lambda\sin\Psi\left(\cos\Psi-\zeta\tau\sin\Psi\right)\right],\label{eq:load:T:lin}
\end{eqnarray}
where
\begin{eqnarray}
J_{4}&=&I_{4}-1=\lambda^2\cos^2\Phi+2\lambda^2\zeta\tau\cos\Phi\sin\Phi+\zeta^2\sin^2\Phi\left(\lambda^2\tau^2+1\right)-1,\label{eq:J4}\\
J_{6}&=&I_{6}-1=\lambda^2\cos^2\Psi-2\lambda^2\zeta\tau\cos\Psi\sin\Psi+\zeta^2\sin^2\Psi\left(\lambda^2\tau^2+1\right)-1.\label{eq:J6}
\end{eqnarray}

Next, we define the following Jacobian matrix \cite{Goriely:2013:GT}
\begin{equation}\label{eq:J}
\textbf{J}=(J_{ij})_{i,j=1,2,3}=\left[
\begin{array}{ccc}
\partial P/\partial\lambda & \partial P/\partial\zeta & \partial P/\partial\tau\\
\partial F/\partial\lambda & \partial F/\partial\zeta & \partial F/\partial\tau\\
\partial T/\partial\lambda & \partial T/\partial\zeta & \partial T/\partial\tau
\end{array}
\right],
\end{equation}
and concentrate our attention on infinitesimal deformations near the reference configuration, with $(\lambda,\zeta,\tau)=(1,1,0)$, where $\textbf{J}|_{(1,1,0)}$ has the following components,
\begin{equation}\label{eq:J:entries}
\begin{split}
J_{11}|_{(1,1,0)}&=\frac{\partial P}{\partial\lambda}|_{(1,1,0)}=4\mu+2\mu_{4}\cos^4\Phi+2\mu_{6}\cos^4\Psi,\\
J_{12}|_{(1,1,0)}&=\frac{\partial P}{\partial\zeta}|_{(1,1,0)}=2\mu+2\mu_{4}\cos^2\Phi\sin^2\Phi+2\mu_{6}\cos^2\Psi\sin^2\Psi,\\
J_{13}|_{(1,1,0)}&=\frac{\partial P}{\partial\tau}|_{(1,1,0)}=2\mu_{4}\cos^3\Phi\sin\Phi-2\mu_{6}\cos^3\Psi\sin\Psi,\\
J_{21}|_{(1,1,0)}&=\frac{\partial F}{\partial\lambda}|_{(1,1,0)}=2\pi\mu_{4}\cos^2\Phi\left(3\sin^2\Phi-1\right)+2\pi\mu_{6}\cos^2\Psi\left(3\sin^2\Psi-1\right),\\
J_{22}|_{(1,1,0)}&=\frac{\partial F}{\partial\zeta}|_{(1,1,0)}=6\pi\mu+2\pi\mu_{4}\sin^2\Phi\left(3\sin^2\Phi-1\right)+2\pi\mu_{6}\sin^2\Psi\left(3\sin^2\Psi-1\right),\\
J_{23}|_{(1,1,0)}&=\frac{\partial F}{\partial\tau}|_{(1,1,0)}=2\pi\mu_{4}\cos\Phi\sin\Phi\left(3\sin^2\Phi-1\right)-2\pi\mu_{6}\cos\Psi\sin\Psi\left(3\sin^2\Psi-1\right),\\
J_{31}|_{(1,1,0)}&=\frac{\partial T}{\partial\lambda}|_{(1,1,0)}=4\pi\mu_{4}\cos^3\Phi\sin\Phi-4\pi\mu_{6}\cos^3\Psi\sin\Psi,\\
J_{32}|_{(1,1,0)}&=\frac{\partial T}{\partial\zeta}|_{(1,1,0)}=4\pi\mu_{4}\cos\Phi\sin^3\Phi-4\pi\mu_{6}\cos\Psi\sin^3\Psi,\\
J_{33}|_{(1,1,0)}&=\frac{\partial T}{\partial\tau}|_{(1,1,0)}=2\pi\mu+4\pi\mu_{4}\cos^2\Phi\sin^2\Phi+4\pi\mu_{6}\cos^2\Psi\sin^2\Psi.
\end{split}
\end{equation}
Assuming that $\det\textbf{J}|_{(1,1,0)} \neq 0$, we can define the matrix inverse
\begin{equation}\label{eq:A}
\textbf{A}=(A_{ij})_{i,j=1,2,3}=\textbf{J}^{-1}|_{(1,1,0)}.
\end{equation}
This will be useful when exploring the critical points where an inversion in the deformation occurs.

\begin{figure}[htbp]
	\begin{center}
		\includegraphics[width=1\textwidth]{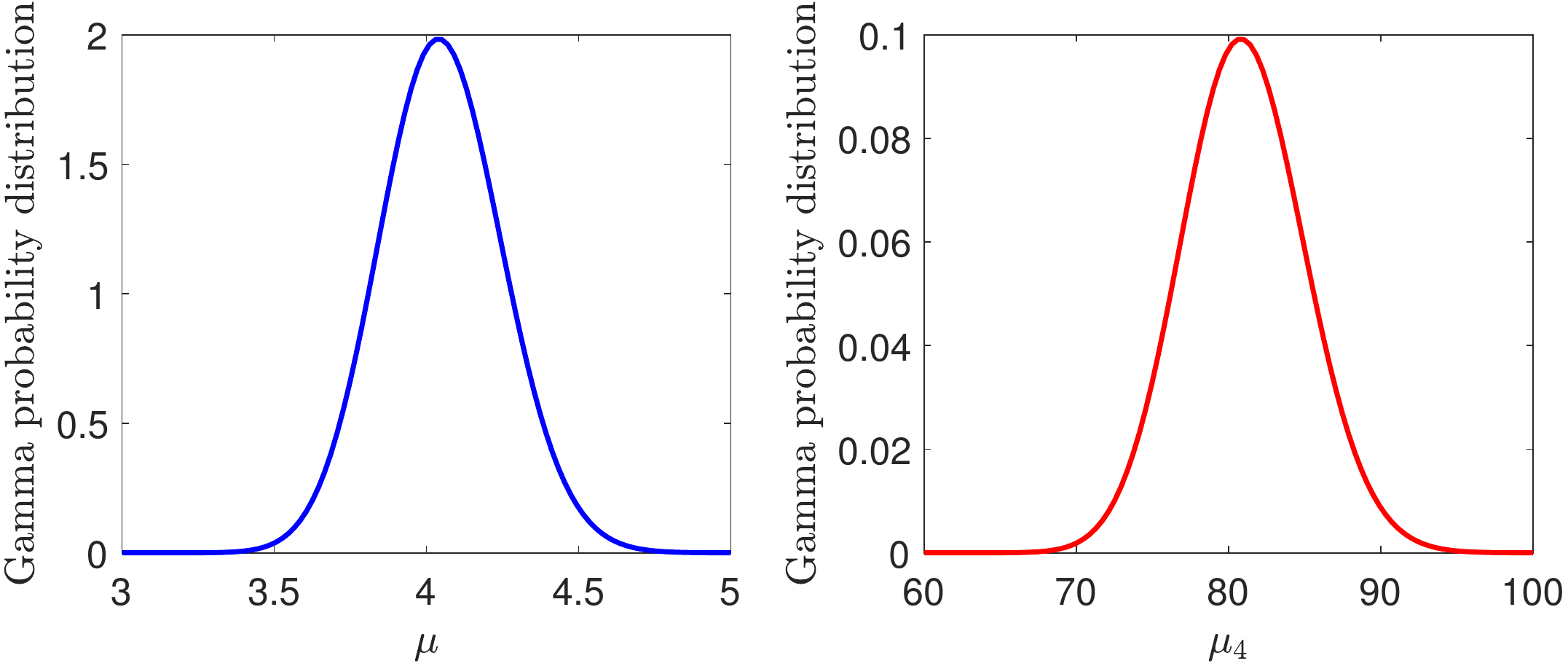}
		\caption{Examples of Gamma distribution with hyperparameters $\rho_{1}=405$ and $\rho_{2}=0.01$ for $\mu>0$ (left), and $\rho_{1}^{(4)}=405$ and $\rho_{2}^{(4)}=0.2$, for $\mu_{4}=\mu_{6}>0$ (right).}\label{fig:mu4mu-gpdfs}
	\end{center}
\end{figure}

In order to obtain clear explicit results that will show the role played by the stochastic parameters, henceforth, we limit our investigation to the case where the anisotropic material has the same mechanical properties in the two preferred directions, i.e., $\mu_{4}=\mu_{6}$. Recalling that $\mu$ follows a Gamma probability distribution $g(u;\rho_{1},\rho_{2})$, defined by \eqref{eq:mu:gamma}, and $\mu_{4}$ follows a Gamma distribution, $g_{4}\left(u;\rho^{(4)}_{1},\rho^{(4)}_{2}\right)$, the probability distribution of each entry of the Jacobian matrix \eqref{eq:J} can be computed using the summation formula for independent Gamma-distributed random variables \cite{Mathai:1982,Moschopoulos:1985} (see Theorem~\ref{th:gammasum} in Appendix~\ref{sec:append}).

To illustrate our theoretical results, numerical examples of the Gamma distributions considered here are represented in Figure~\ref{fig:mu4mu-gpdfs}, where the shape and scale parameters are $\rho_{1}=405$ and $\rho_{2}=0.01$ for $\mu>0$, and $\rho_{1}^{(4)}=\rho_{1}=405$ and $\rho_{2}^{(4)}=20\rho_{2}=0.2$ for $\mu_{4}=\mu_{6}=20\mu>0$. Note that, although Gamma distributions are used throughout this paper, those represented in Figure~\ref{fig:mu4mu-gpdfs} appear to be approximately normal distributions. This is not a coincidence, and we discussed such a comparison in \cite{Mihai:2018b:MDWG}, in the case when $\rho_{1}$ was large compared to $\rho_{2}$. However, despite known convergence results and the qualitative agreement between the two density functions for large values of the mean, in general, the normal distribution cannot be used to model elastic parameters. This is due to the fact that the normal distribution is defined on the entire real line, whereas elastic moduli are always positive. Approaches for the explicit derivation of probability distributions for the constitutive parameters of stochastic homogeneous isotropic hyperelastic models, calibrated to experimental data, were developed in \cite{Mihai:2018:MWG,Staber:2017:SG}. As far as we are aware, similar derivations for the stochastic constitutive laws of anisotropic materials are yet to be performed. In practice, elastic model parameters can meaningfully take on different values, corresponding to possible outcomes of the experimental tests. Then, by the maximum entropy principle, one can explicitly construct probability laws using the available information.

\subsection{Likely inflation of the stochastic tube}

First, we assume that $\Phi=\Psi$ \cite{Goriely:2013:GT}, and consider the critical point for radial stretch $\lambda$, such that
\begin{equation}\label{eq:A11}
A_{11}=0,\qquad\mbox{where}\qquad A_{11}=\frac{\mu_{4}\left[3\cos^2(2\phi)-4\cos(2\phi)+1\right]+6\mu}{8\mu\left[3\mu_{4}\cos^2(2\phi)+\mu_{4}+3\mu\right]}.
\end{equation}
The radial stretch increases if $A_{11}>0$ and decreases if $A_{11}<0$, and equation \eqref{eq:A11} is equivalent to the following quadratic equation in $\cos(2\Phi)$,
\begin{equation}\label{eq:radial:inv}
\mu_{4}\left[3\cos^2(2\Phi)-4\cos(2\Phi)+1\right]+6\mu=0,
\end{equation}
which has real solutions when $0<\mu\leq\mu_{4}/18$. As the denominator of $A_{11}$ is positive, it follows that, as the internal pressure increases, the radial stretch, $\lambda$, increases if $\mu>\mu_{4}/18$, and decreases if $0<\mu<\mu_{4}/18$.

\begin{figure}[htbp]
	\begin{center}
		\includegraphics[width=1\textwidth]{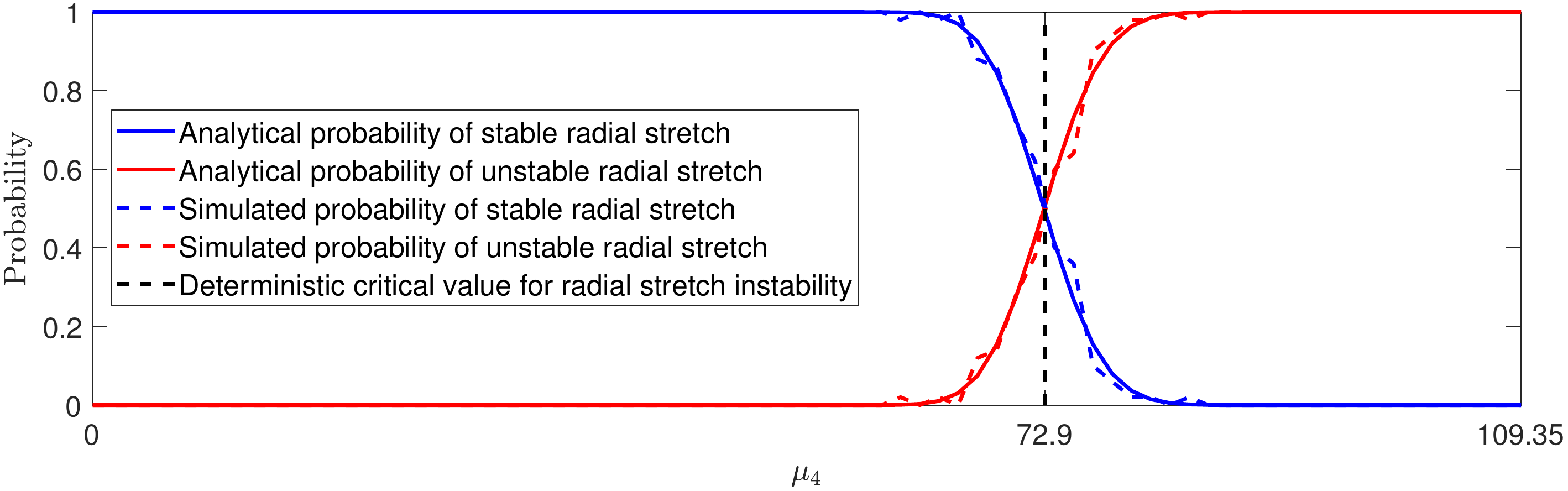}
		\caption{Probability distributions of whether radial stretch instability can occur or not for an anisotropic cylindrical tube of stochastic hyperelastic material described by \eqref{eq:W:stoch}, when the shear modulus, $\mu$, follows a Gamma distribution with $\rho_{1}=405$, $\rho_{2}=0.01$, and $\mu_{4}$ follows a Gamma distribution with $\rho_{1}^{(4)}=405$, $\rho_{2}^{(4)}=0.2$. Continuous coloured lines represent analytically derived solutions, given by equations \eqref{eq:P1}-\eqref{eq:P2}, whereas the dashed versions represent stochastically generated data comprised of 100 simulations. The vertical line at the critical value, $\mu_{4}=72.9$, separates the expected regions based only on mean parameter value, $\underline{\mu}=\rho_{1}\rho_{2}=4.05$.}\label{fig:a11-intpdfs}
	\end{center}
	\begin{center}
		\includegraphics[width=.5\textwidth]{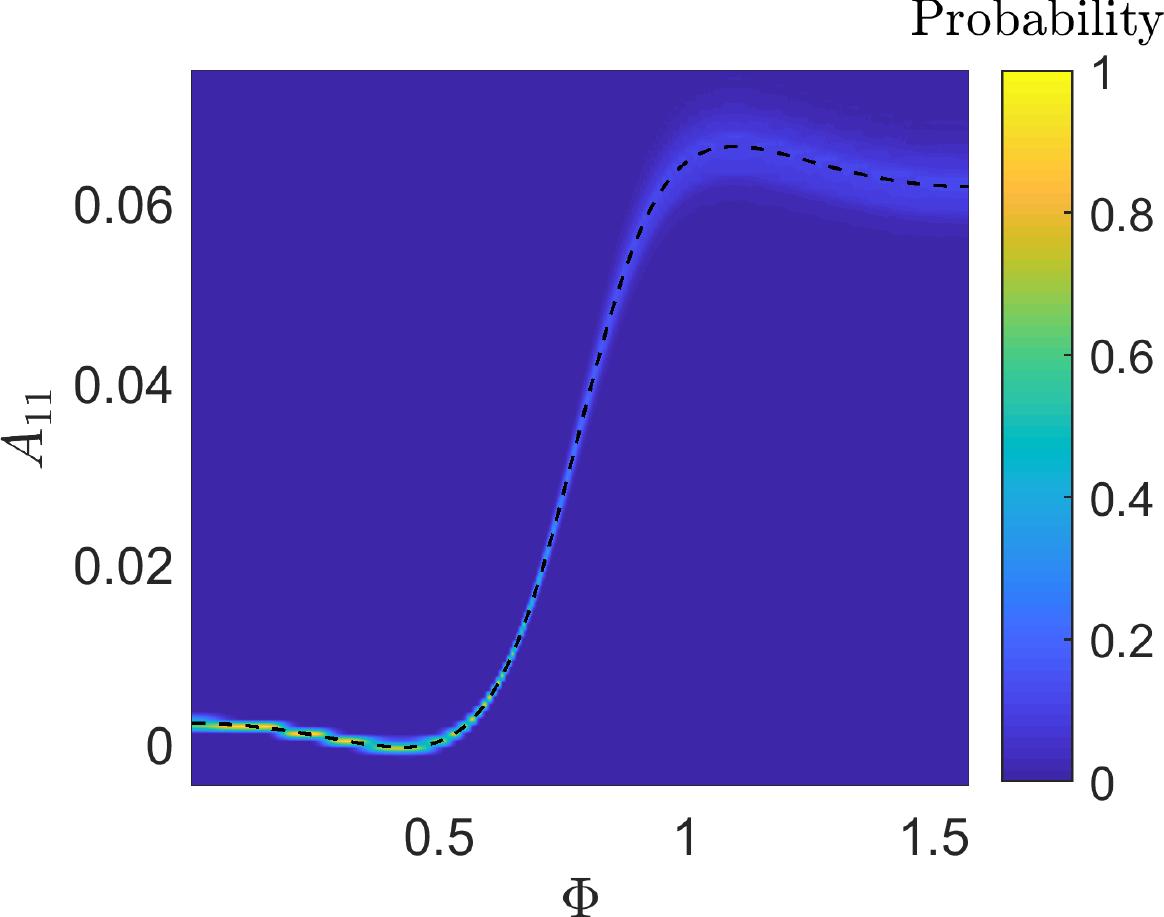}
		\caption{Probability distribution of stochastic $A_{11}$, for the inflation of an anisotropic cylindrical tube of stochastic hyperelastic material, given by \eqref{eq:W:stoch}, where the shear modulus, $\mu$, is drawn from a Gamma distribution with $\rho_{1}=405$, $\rho_{2}=0.01$, and $\mu_{4}$ is drawn from a Gamma distribution with $\rho_{1}^{(4)}=405$, $\rho_{2}^{(4)}=0.2$. As $\underline{\mu}=\rho_{1}\rho_{2}=4.05<\underline{\mu}_{4}/18=\rho_{1}^{(4)}\rho_{2}^{(4)}/18=4.5$, instability is expected to occur, but there is also around 2\% chance that the radial stretch is stable. The dashed black line corresponds to the expected value of $A_{11}$ based only on mean parameter values, $\underline{\mu}=\rho_{1}\rho_{2}=4.05$ and $\underline{\mu}_{4}=\rho_{1}^{(4)}\rho_{2}^{(4)}=81$. Each distribution was calculated from the average of $1000$ stochastic simulations.}\label{fig:a11-stochastic}
	\end{center}
\end{figure}

\begin{figure}[htbp]
	\begin{center}
		\includegraphics[width=1\textwidth]{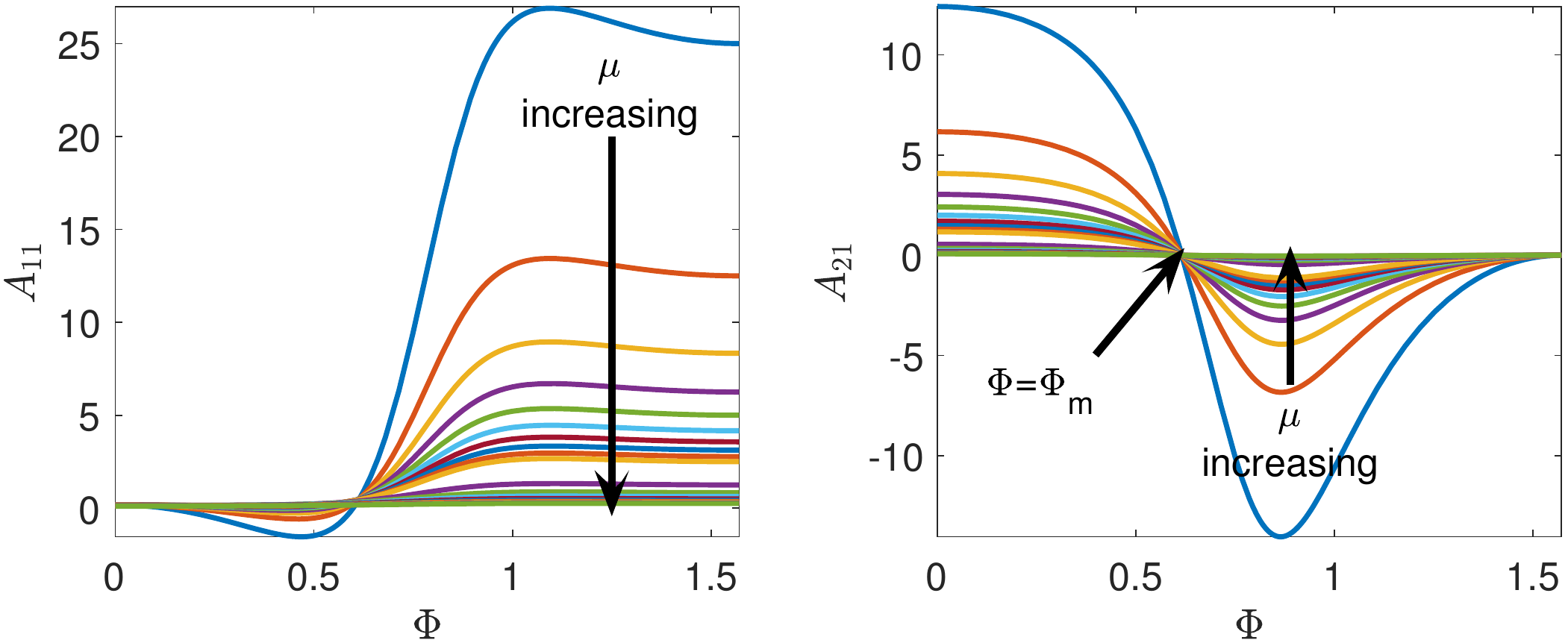}
		\caption{Deterministic values of $A_{11}$ (left) and $A_{21}$ (right), for $\mu_{4}=1$ and $\mu\in\{0.01,0.02, \cdots, 0.1,0.2,\cdots,1\}$ (the direction of increasing values of $\mu$ is indicated by arrow).}\label{fig:a11a21}
	\end{center}
\end{figure}

In this case, we can express the probability distribution of stable inflation, such that the radial stretch monotonically increases when the internal pressure increases, as
\begin{equation}\label{eq:P1}
P_{1}(\mu_{4})=1-\int_{0}^{\mu_{4}/18}g(u;\rho_{1},\rho_{2})du,
\end{equation}
and that of unstable inflation, whereby the radial stretch starts to decreases under increasing pressure, as
\begin{equation}\label{eq:P2}
P_{2}(\mu_{4})=1-P_{1}(\mu_{4}).
\end{equation}

For example, taking $\rho_{1}=405$ and  $\rho_{2}= 0.01$ (see Figure~\ref{fig:mu4mu-gpdfs}-left), the mean value of the shear modulus is $\underline{\mu}=\rho_{1}\rho_{2}=4.05$, and the probability distributions given by equations \eqref{eq:P1}-\eqref{eq:P2} are illustrated in Figure~\ref{fig:a11-intpdfs} (blue lines for $P_{1}$ and red lines for $P_{2}$). To plot those results, the interval $\left(0,27\underline{\mu}\right)$ was discretised into $100$ representative points, then for each value of $\mu_{4}$, $100$ random values of $\mu$ were numerically generated from the specified Gamma distribution and compared with the inequalities defining the two intervals for values of $\mu_{4}$. For the deterministic elastic case, which is based on the mean value of the shear modulus, $\underline{\mu}=\rho_{1}\rho_{2}=4.05$, the critical value of $\mu_{4}=18\mu=72.9$ strictly separates the cases where radial stretch instability can, and cannot, occur. For the stochastic problem, for the same critical value, there is, by definition, exactly 50\% chance that radial stretch is stable (blue lines) and 50\% chance that instability occurs (red lines). To increase the probability of stable radial stretch ($P_{1}\approx 1$), one must consider values of $\mu_{4}$ that are sufficiently smaller than the expected critical value, whereas an instability is almost certain to occur ($P_{2}\approx 1$) for values of $\mu_{4}$ which are sufficiently greater than the expected critical value. However, the inherent variability in the probabilistic system means that there exist events where there is competition between the two cases.

In Figure~\ref{fig:a11-stochastic}, we represent the probability distribution of $A_{11}$ as a function of the angle $\Phi$ when $\mu$ follows a Gamma distribution with hyperparameters $\rho_{1}=405$ and $\rho_{2}=0.01$, and $\mu_{4}$ follows a Gamma distribution with $\rho_{1}^{(4)}=405$ and $\rho_{2}^{(4)}=0.2$ (see Figure~\ref{fig:mu4mu-gpdfs}). In this case, $\underline{\mu}_{4}=20\underline{\mu}<81$, and unstable radial stretch is expected. Nevertheless, the probability distribution suggests that there is also around 2\% chance that the radial stretch is stable. 

For the axial stretch $\zeta$, the critical value satisfies the equation
\begin{equation}\label{eq:A21}
A_{21}=0\qquad\mbox{where}\qquad
A_{21}=\frac{\mu_{4}\left[3\cos^2(2\phi)+2\cos(2\phi)-1\right]}{8\mu\left[3\mu_{4}\cos^2(2\phi)+\mu_{4}+3\mu\right]},
\end{equation}
and the axial stretch increases if $A_{21}>0$, and decreases if $A_{21}<0$. Solving equation \eqref{eq:A21}, we obtain
\begin{equation}\label{eq:magic}
\Phi=\Phi_{m}=\frac{1}{2}\arccos\frac{1}{3}\approx35.3^{\circ},
\end{equation}
where $\Phi_{m}$ denotes the `magic angle' \cite[p.~337]{goriely17}. In this case, as the denominator of $A_{21}$ is always positive and \eqref{eq:magic} is independent of the random material parameters $\mu$ and $\mu_{4}$, there is no uncertainty to be resolved regarding this instability. For the deterministic cases where $\mu_{4}=1$ and $\mu\in\{0.01,0.02, \cdots, 0.1,0.2,\cdots,1\}$, the values of $A_{11}$ and $A_{21}$ are presented graphically in Figure~\ref{fig:a11a21}.

\subsection{Likely chirality of the stochastic tube}

Next, we assume that the angle of a preferred direction is kept fixed, while the other angle can vary \cite{Goriely:2013:GT}. When the radius and axial length of the tube do not change, the condition that the torsion parameter $\tau$ may start to decrease when the internal pressure increases is
\begin{equation}\label{eq:A31}
A_{31}=0.
\end{equation}
This is equivalent to
\begin{equation}\label{eq:angular:inv}
\begin{split}
&6\mu\left[2\sin(2\Phi)+\sin(4\Phi)-2\sin(2\Psi)-\sin(4\Psi)\right]\\
&=\mu_{4}\left[3\sin(2\Psi)+\sin(4\Psi)-3\sin(2\Phi)-\sin(4\Phi)\right.
\left.+2\sin(2\Phi-2\Psi)-3\sin(2\Phi+4\Psi)+3\sin(4\Phi+2\Psi)\right],
\end{split}
\end{equation}
which is clearly satisfied when $\Phi=\Psi$, or if $\Phi\in\{0,\pi/2\}$ and $\Psi\in\{0,\pi/2\}$.

\begin{figure}[htbp]
	\begin{center}
		\includegraphics[width=.7\textwidth]{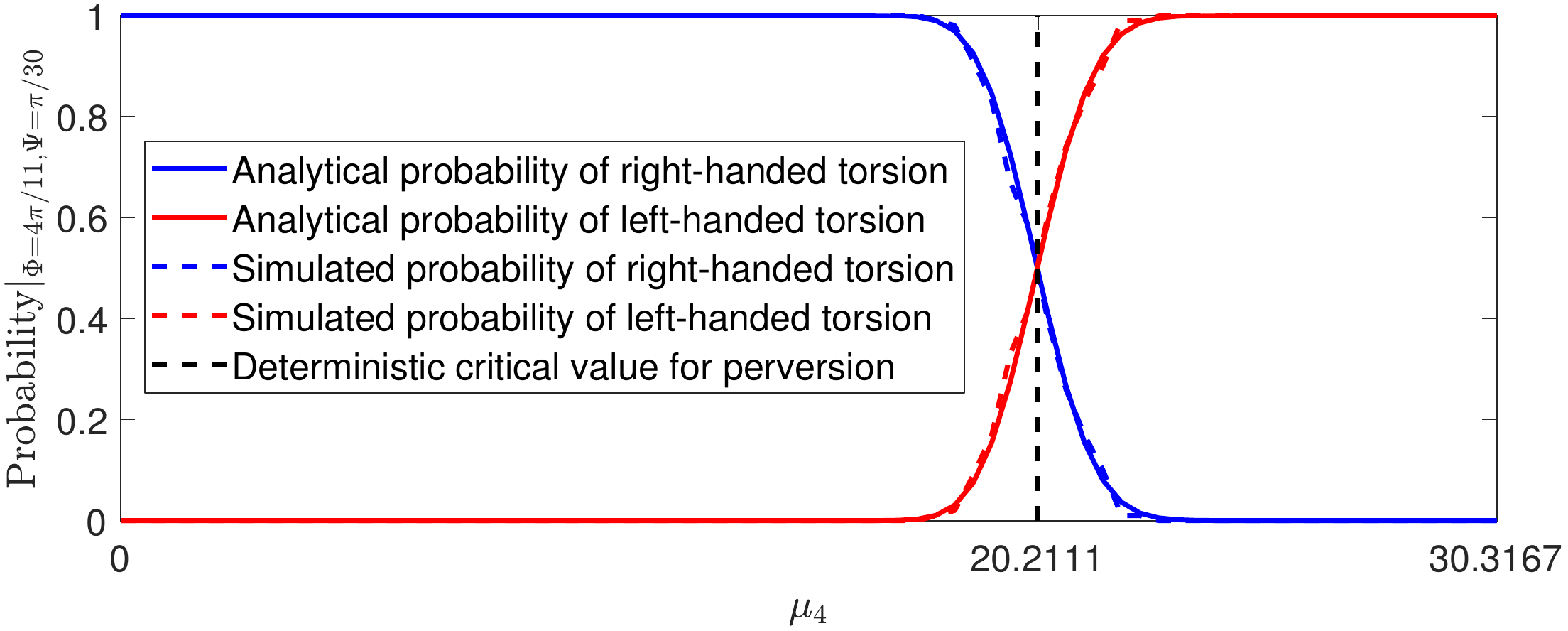}\\
		\includegraphics[width=.7\textwidth]{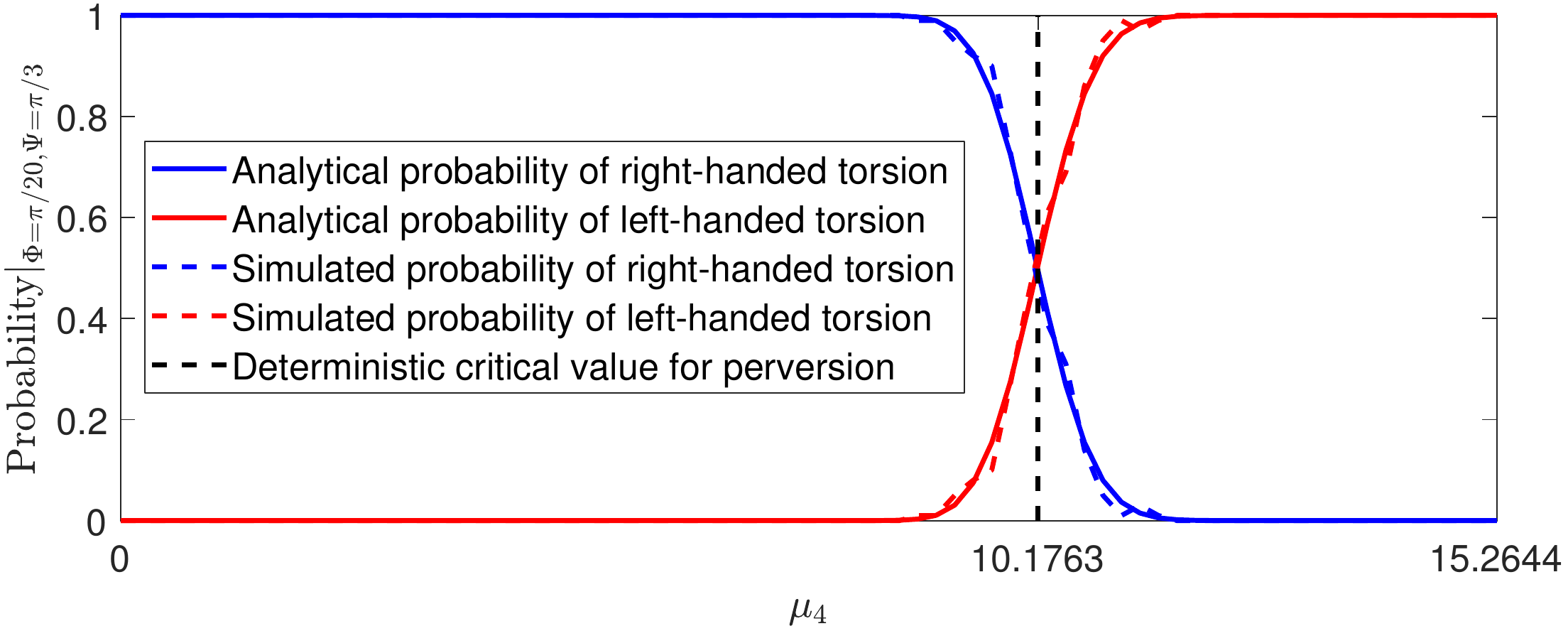}\\
		\includegraphics[width=.7\textwidth]{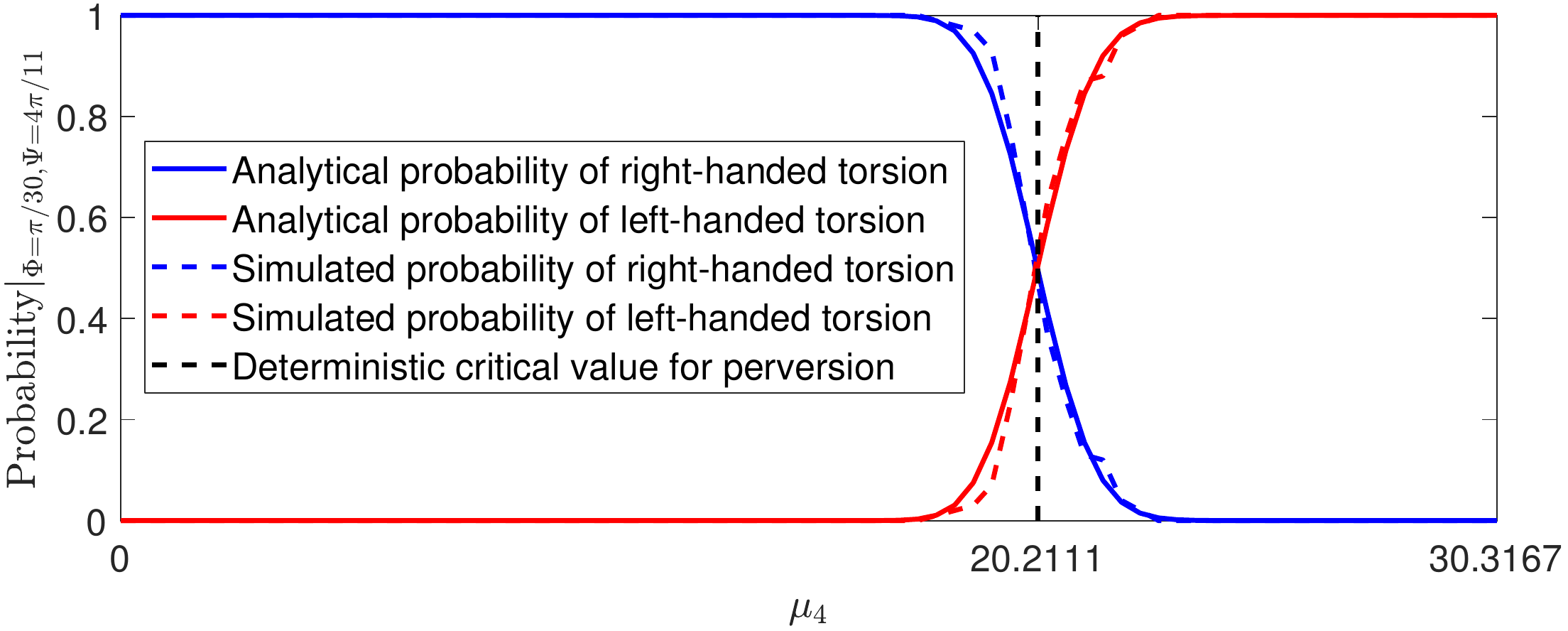}\\
		\includegraphics[width=.7\textwidth]{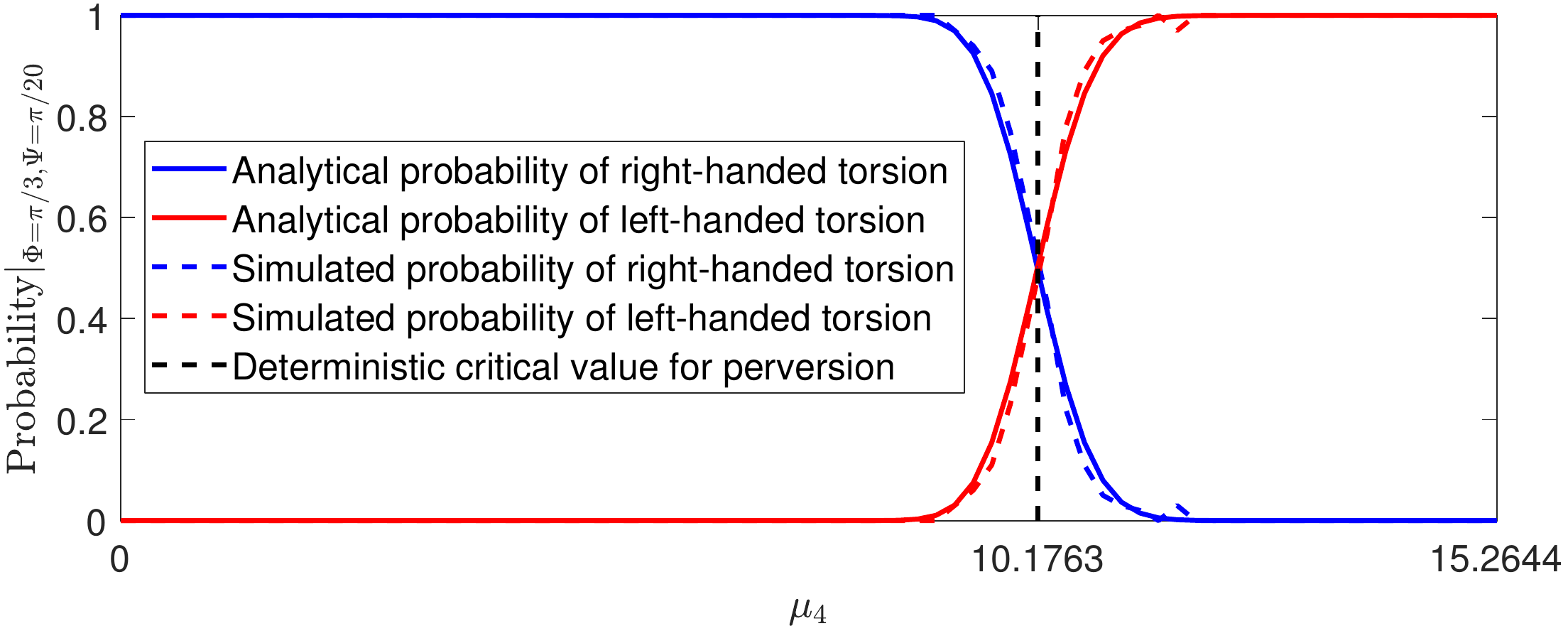}
		\caption{Probability distributions of whether perversion can occur or not for an anisotropic cylindrical tube of stochastic hyperelastic material described by \eqref{eq:W:stoch}, when the shear modulus, $\mu$, follows a Gamma distribution with $\rho_{1}=405$, $\rho_{2}=0.01$, and $\mu_{4}$ follows a Gamma distribution with $\rho_{1}^{(4)}=405$, $\rho_{2}^{(4)}=0.2$, for the cases where (from top to bottom): $\Phi=4\pi/11$ and $\Psi=\pi/30$; $\Phi=\pi/20$ and $\Psi=\pi/3$; $\Phi=\pi/30$ and $\Psi=4\pi/11$; $\Phi=\pi/3$ and $\Psi=\pi/20$. Continuous coloured lines represent analytically derived solutions, given by equations \eqref{eq:P3}-\eqref{eq:P4}, while the dashed versions represent stochastically generated data comprised of 100 simulations. The vertical line separates the expected regions based only on mean parameter value, $\underline{\mu}=\rho_{1}\rho_{2}=4.05$.}\label{fig:a31-intpdfs}
	\end{center}
\end{figure}

\begin{figure}[htbp]
	\begin{center}
    \includegraphics[width=.45\textwidth]{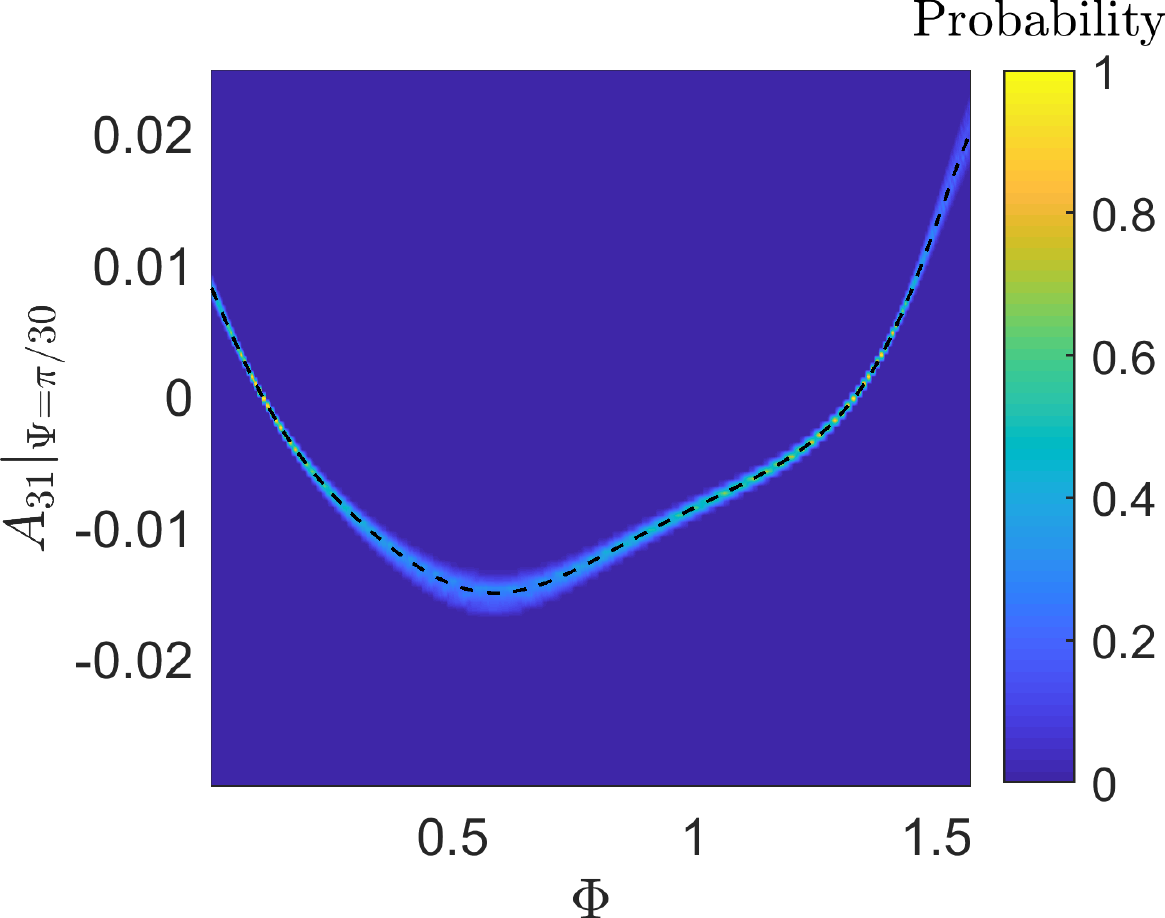}
    \includegraphics[width=.45\textwidth]{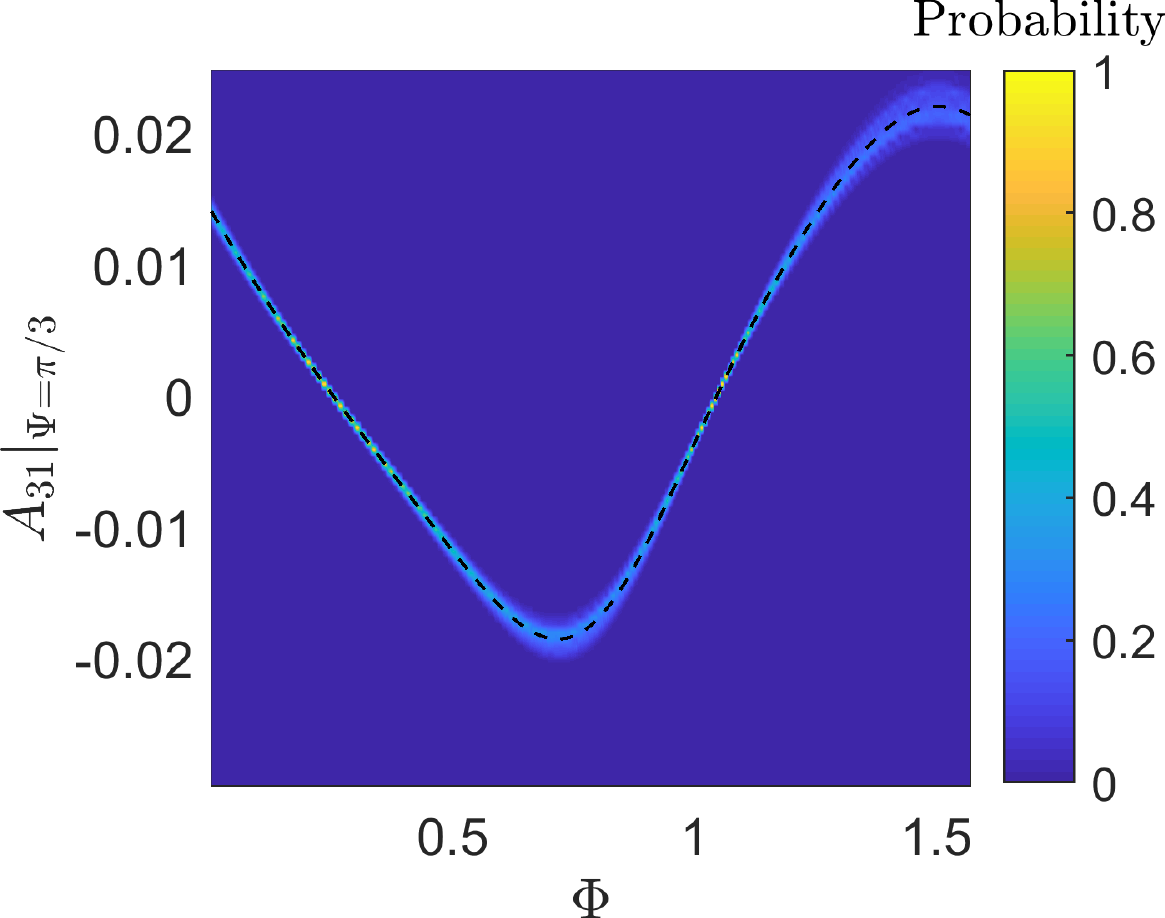}
	\includegraphics[width=.45\textwidth]{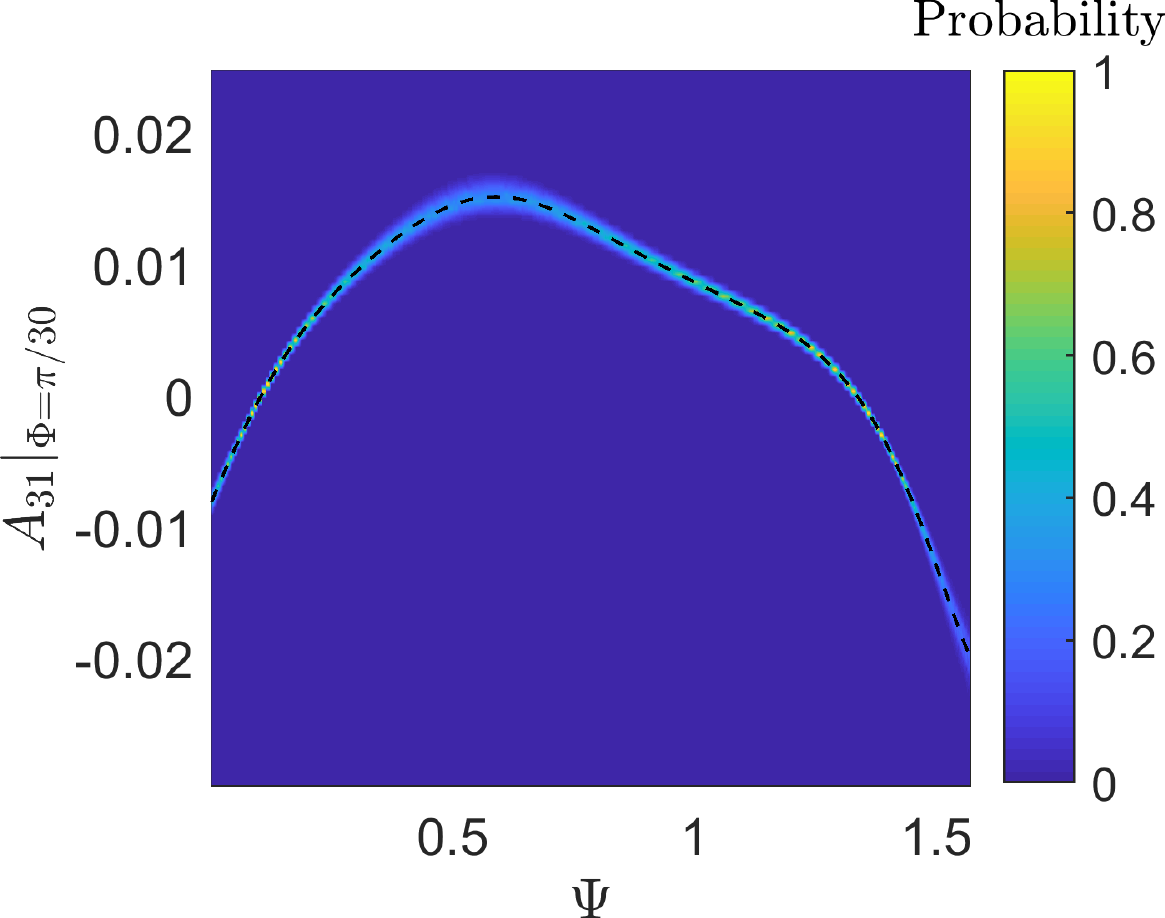}
	\includegraphics[width=.45\textwidth]{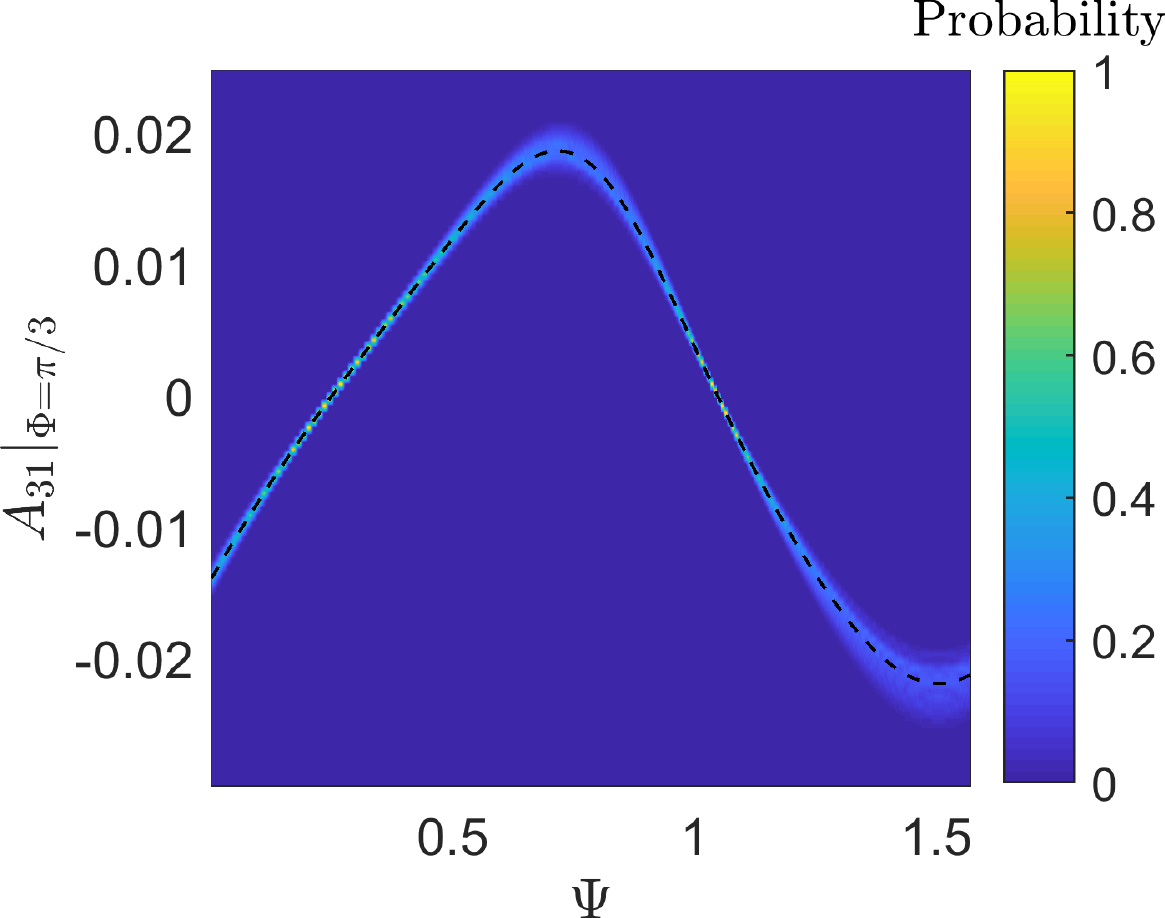}	
	\caption{Probability distribution of stochastic $A_{31}$ for the torsion of an anisotropic cylindrical tube of stochastic hyperelastic material, given by \eqref{eq:W:stoch}, when $\mu$ is drawn from a Gamma distribution with $\rho_{1}=405$, $\rho_{2}=0.01$, and $\mu_{4}$ is drawn from a Gamma distribution with $\rho_{1}^{(4)}=405$, $\rho_{2}^{(4)}=0.2$, for the cases where: $\Psi_{0}=\pi/30$ (top-left); $\Psi_{0}=\pi/3$ (top-right); $\Phi_{0}=\pi/30$ (bottom-left); $\Phi_{0}=\pi/3$ (bottom-right). In these cases, perversion takes place with certainty at $\Psi=\Phi_{0}$, and  is expected to occur also, with a given probability, for a different angle $\Psi$. The dashed black line corresponds to the expected value of $A_{31}$ based only on mean parameter values, $\underline{\mu}=\rho_{1}\rho_{2}=4.05$ and $\underline{\mu}_{4}=\rho_{1}^{(4)}\rho_{2}^{(4)}=81$.}\label{fig:a31-stochastic}
	\end{center}
\end{figure}

\begin{figure}[htbp]
	\begin{center}
		\includegraphics[width=0.9\textwidth]{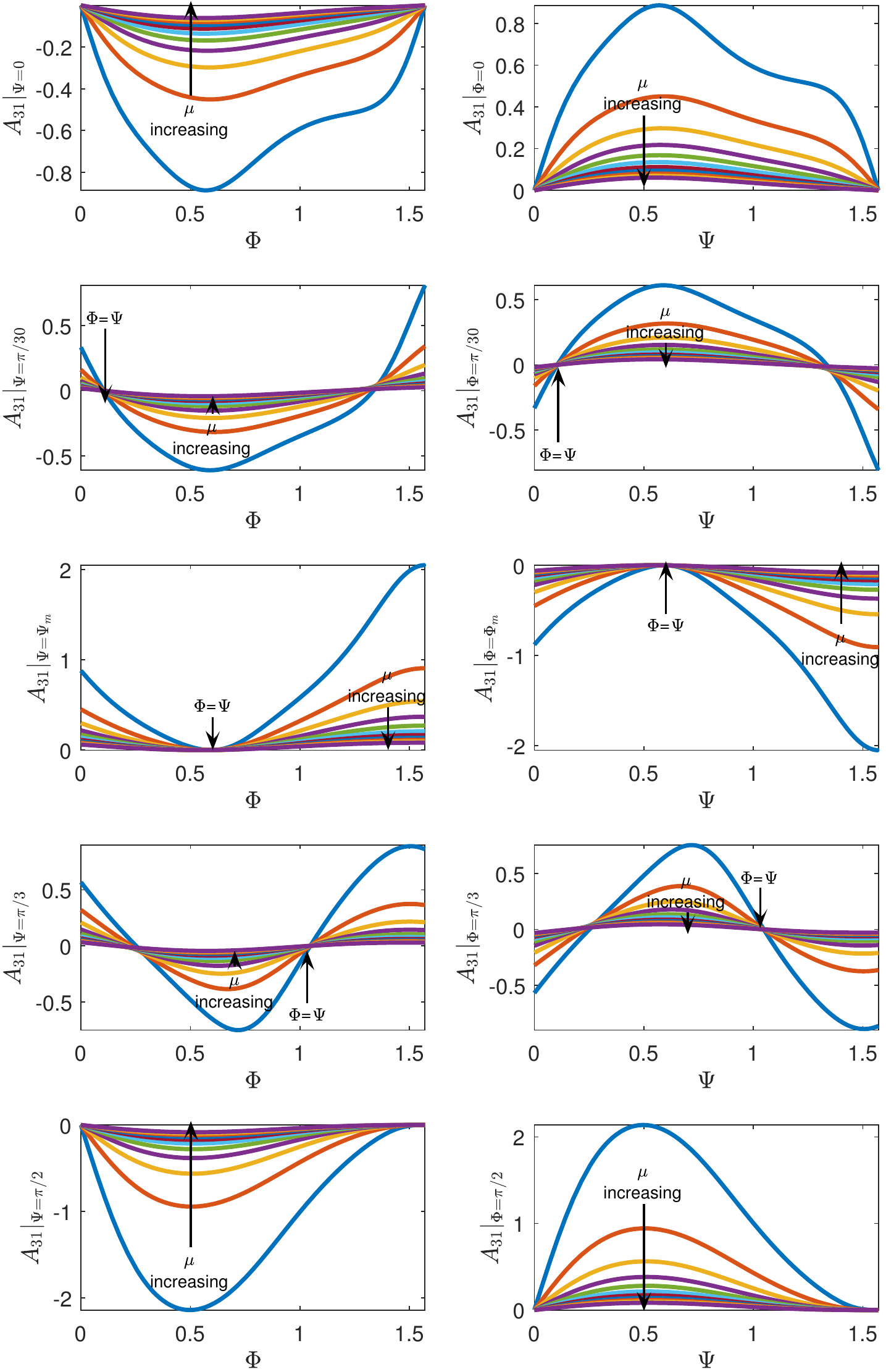}
		\caption{Deterministic values of $A_{31}$ when $\Psi\in\{0,\pi/30,\Psi_{m},\pi/3,\pi/2\}$ is fixed and $\Phi$ varies (left), and when $\Phi\in\{0,\pi/30,\Phi_{m},\pi/3,\pi/2\}$ is fixed and $\Psi$ varies (right), for $\mu_{4}=1$ and $\mu\in\{0.1,0.2,\cdots,1\}$ (the direction of increasing values of $\mu$ is indicated by arrow).}\label{fig:a31}
	\end{center}
\end{figure}

\begin{figure}[htbp]
	\begin{center}
		\includegraphics[width=1\textwidth]{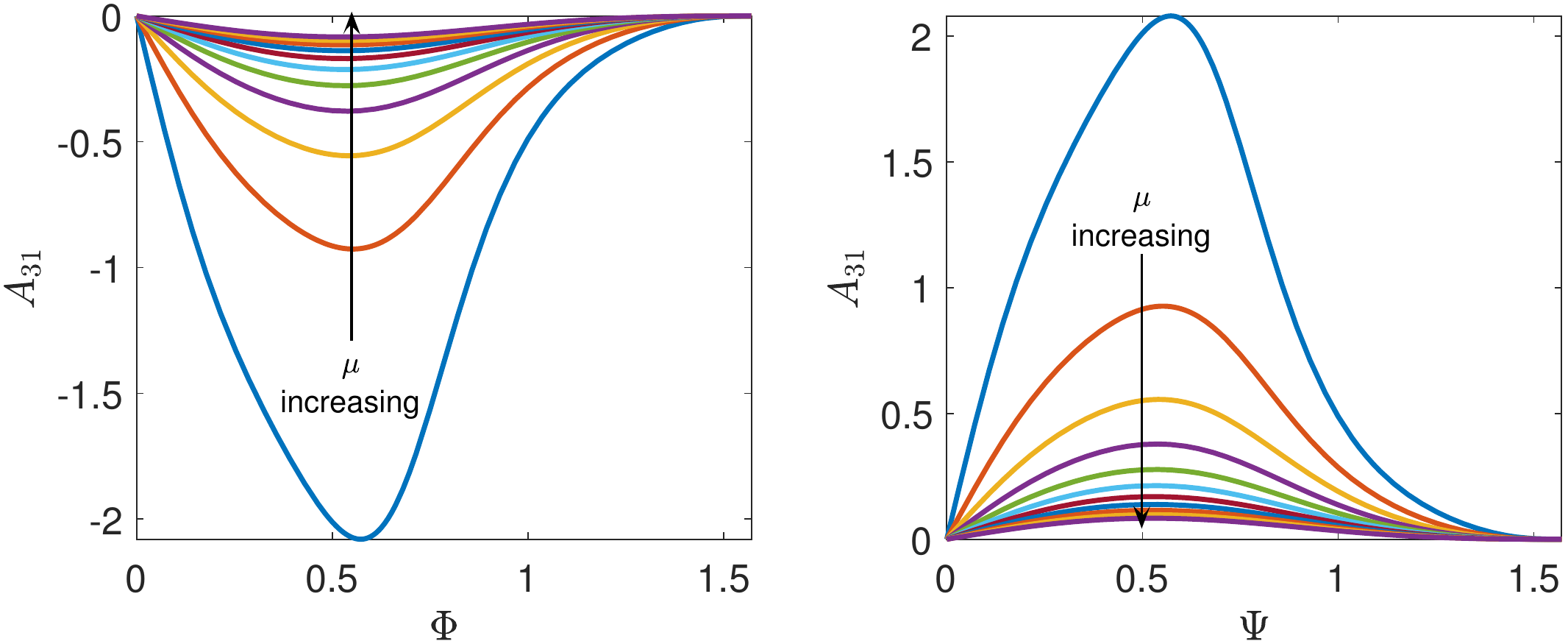}
		\caption{The values of $A_{31}$ in the case of a single family of right-handed fibres (left), or left-handed fibres (right), for $\mu_{4}=1$ and $\mu\in\{0.1,0.2,\cdots,1\}$ (the arrow shows the direction of increasing values of $\mu$).}\label{fig:a31-single}
	\end{center}
\end{figure}

When $\Phi\neq\Psi$ and either $\Phi\notin\{0,\pi/2\}$ or $\Psi\notin\{0,\pi/2\}$, we define
\begin{equation}\label{eq:xi4}
\xi=\frac{3\sin(2\Psi)+\sin(4\Psi)-3\sin(2\Phi)-\sin(4\Phi)+2\sin(2\Phi-2\Psi)-3\sin(2\Phi+4\Psi)+3\sin(4\Phi+2\Psi)}{6\left[2\sin(2\Phi)+\sin(4\Phi)-2\sin(2\Psi)-\sin(4\Psi)\right]},
\end{equation}
and distinguish the following cases:
\begin{itemize}
\item[($i$)] When $\Psi=\Psi_{0}$ is fixed and $\Phi\in(0,\pi/2)\setminus\{\Psi_{0}\}$:
    \begin{itemize}
	\item[($i_1$)] if $\Psi_{0}\in\{0,\pi/2\}$, then $A_{31}<0$, i.e., there is left-handed (clockwise) twist;
	\item[($i_2$)] if $\Psi_{0}=\Psi_{m}$, where $\Psi_{m}$ is the magic angle given by \eqref{eq:magic}, then $A_{31}>0$, i.e., there is right-handed (anti-clockwise) twist;
	\item[($i_3$)] if $0<\Psi_{0}<\Psi_{m}$, then $A_{31}>0$ (right-handed twist) when $\Phi<\Psi_{0}$, and there exists $\Phi^{*}>\Psi_{0}$, such that $A_{31}<0$  (left-handed twist) when $\Psi_{0}<\Phi<\Phi^{*}$ and $A_{31}>0$  (right-handed twist) when $\Phi^{*}<\Psi<\pi/2$. Hence, a first inversion from right-handed to left-handed twist occurs at $\Phi=\Psi_{0}$, and a second one from left-handed to right-handed twist takes place at $\Phi=\Phi^{*}$.

	Then, if $\Psi_{0}<\Phi<\pi/2$, the torsion parameter $\tau$ increases (i.e., $A_{31}>0$) when $\mu>\mu_{4}/\xi$ and decreases when $0<\mu<\mu_{4}/\xi$, and if $\Phi<\Psi$, the torsion parameter increases when $0<\mu<\mu_{4}/\xi$ and decreases (i.e., $A_{31}<0$) when $\mu>\mu_{4}/\xi$. Thus, the probability distribution of right-handed torsion, such that $\tau$ monotonically increases as the internal pressure increases, is
	\begin{equation}\label{eq:P3}
	P_{3}(\mu_{4})=1-\int_{0}^{\mu_{4}/\xi}g(u;\rho_{1},\rho_{2})du,
	\end{equation}
	and that of left-handed torsion, such that $\tau$ decreases as the pressure increases, is
	\begin{equation}\label{eq:P4}
	P_{4}(\mu_{4})=1-P_{3}(\mu_{4}).
	\end{equation}
	
	\item[($i_4$)] if $\Psi_{m}<\Psi_{0}<\pi/2$, then $A_{31}>0$ (right-handed twist) when $\Phi>\Psi_{0}$, and there exists $\Phi^{*}<\Psi_{0}$, such that $A_{31}>0$  (right-handed twist) when $0<\Phi<\Phi^{*}$ and $A_{31}<0$  (left-handed twist) when $\Phi^{*}<\Psi<\Psi_{0}$. In this case, the first inversion from right-handed to left-handed twist occurs at $\Phi=\Phi^{*}$, and the second one from left-handed to right-handed twist at $\Phi=\Psi_{0}$.
	
	Hence, for $0<\Phi<\Psi_{0}$, the probability of right-handed torsion is $P_{4}(\mu_{4})$, given by \eqref{eq:P4}, and that of left-handed torsion is $P_{3}(\mu_{4})$, given by \eqref{eq:P3}.
    \end{itemize}
\item[($ii$)] When $\Phi=\Phi_{0}$ is fixed and $\Psi\in(0,\pi/2)\setminus\{\Phi_{0}\}$:
    \begin{itemize}
	\item[($ii_1$)] if $\Phi_{0}\in\{0,\pi/2\}$, then $A_{31}>0$, i.e., there is right-handed twist;
	\item[($ii_2$)] if $\Phi_{0}=\Phi_{m}$, where $\Phi_{m}$ is the magic angle given by \eqref{eq:magic}, then $A_{31}<0$, i.e., there is left-handed twist;
	\item[($ii_3$)] if $0<\Phi_{0}<\Phi_{m}$, then  $A_{31}<0$ (left-handed twist) when $0<\Psi<\Phi_{0}$, and there exists $\Psi^{*}>\Phi_{0}$, such that $A_{31}>0$ (right-handed twist) when $\Phi_{0}<\Psi<\Psi^{*}$ and $A_{31}<0$ (left-handed twist) when $\Psi^{*}<\Psi<\pi/2$. Hence, a first inversion from left-handed to right-handed twist takes place at $\Psi=\Phi_{0}$, and a second one from right-handed to left-handed twist at $\Psi=\Psi^{*}$.
	
	Then, for $\Psi_{0}<\Phi<\pi/2$, the probability of right-handed torsion is $P_{4}(\mu_{4})$, given by \eqref{eq:P4}, and that of left-handed torsion is $P_{3}(\mu_{4})$, given by \eqref{eq:P3}.
	
	\item[($ii_4$)] if $\Phi_{m}<\Phi_{0}<\pi/2$, then $A_{31}<0$ (left-handed twist) when $\Phi_{0}<\Psi<\pi/2$, and there exists $\Psi^{*}<\Phi_{0}$, such that $A_{31}<0$ (left-handed twist) when $0<\Psi<\Psi^{*}$ and $A_{31}>0$ (right-handed twist) when $\Psi^{*}<\Psi<\Phi_{0}$. In this case, the first inversion from left-handed to right-handed twist occurs at $\Psi=\Psi^{*}$, and the second one from left-handed to right-handed twist at $\Psi=\Phi_{0}$.
	
	Hence, for $0<\Phi<\Psi_{0}$, the probability of right-handed torsion is $P_{3}(\mu_{4})$, given by \eqref{eq:P3}, and that of left-handed torsion is $P_{4}(\mu_{4})$, given by \eqref{eq:P4}.
	\end{itemize}
\end{itemize}

For example, when $\rho_{1}=405$ and $\rho_{2}= 0.01$ (see Figure~\ref{fig:mu4mu-gpdfs}-left), the probability distributions given by equations \eqref{eq:P3}-\eqref{eq:P4} are illustrated numerically in Figure~\ref{fig:a31-intpdfs}, where the different plots, from top to bottom, correspond, respectively, to the cases: ($i_{3}$) with $\Phi=4\pi/11$ and $\Psi=\pi/30$; ($i_{4}$) with $\Phi=\pi/20$ and $\Psi=\pi/3$; ($ii_{3}$) with $\Phi=\pi/30$ and $\Psi=4\pi/11$; ($ii_{4}$) with $\Phi=\pi/3$ and $\Psi=\pi/20$ (blue lines for $P_{3}$ and red lines for $P_{4}$). In each case, the interval $(0,2\xi\underline{\mu})$ was discretised into $100$ representative points, then for each value of $\mu_{4}$, $100$ random values of $\mu$ were numerically generated from the specified Gamma distribution and compared with the inequalities defining the two intervals for values of $\mu_{4}$. For example, if $\mu_{4}$ follows a Gamma distribution with $\rho_{1}^{(4)}=405$, $\rho_{2}^{(4)}=0.2$ (see Figure~\ref{fig:mu4mu-gpdfs}-right), then perversion will take place with certainty at $\Phi=\Psi$, and is expected to occur also, with a given probability, at a different angle, as seen from Figure~\ref{fig:a31-stochastic}.

For the deterministic cases where $\mu_{4}=1$ and $\mu\in\{0.1,0.2,\cdots,1\}$, the values of $A_{31}$ are illustrated in Figure~\ref{fig:a31}, where $\Psi_{0}\in\{0,\pi/30,\Psi_{m},\pi/3,\pi/2\}$ (left) and $\Phi_{0}\in\{0,\pi/30,\Phi_{m},\pi/3,\pi/2\}$ (right). From the respective plots, we see that a pressurised tube with a right-handed family of fibres deforms by a left-handed torsion if the other fibres are kept either horizontal or vertical (top and bottom left), and a tube with a left-handed family of fibres deforms by a right-handed torsion if the other fibres are horizontal or vertical (top and bottom right). This is the expected behaviour in the case of a single family of fibres as well \cite{Goriely:2013:GT} (see Figure~\ref{fig:a31-single}).

\section{Conclusion}\label{sec:conclude}

We have developed here a procedure for examining the nonlinear elastic responses of a tube of stochastic hyperelastic material with two preferred directions under three typical combined loadings. In our approach, by following the usual finite strain analysis, but with the model parameters treated as random variables defined in terms of probability distributions, instead of taking them as single-valued constants as in the deterministic case, leads to mechanical responses that are also characterised in terms of probability distributions. For a direct comparison of our stochastic results with the deterministic ones, we sampled from distributions where the parameters were set to have mean values corresponding to the deterministic system. Thus, the mean value of the distribution was guaranteed to converge to the expected value. Specifically, we found that, while for the deterministic tube, there is a strict transition between different behaviours, given by a single numerical value, for the stochastic tube, our analysis identifies probabilistic intervals where the two different states compete and both have a quantifiable chance to be presented. 
	
In general, the phenomena of inversion and perversion in these systems are primarily driven by the geometry and fibre distribution. In order to clearly reveal the additional influence of the stochastic material parameters, we considered here the mean directions of two families of aligned fibres, without including the stochastic variations in the fibre angles. The additional influence of the fibre distribution has been studied by various authors \cite{gaogho06,medago15}, who homogenised the effects of stochastic dispersion, which then entered as a parameter in the deterministic setting. Nevertheless, our approach could also be extended to the problem of fibre dispersion to reveal the full effect of stochasticity.

The present study is part of an ongoing investigation where we aim to illustrate explicitly how the elastic solution of fundamental nonlinear elasticity problems can be extended to the stochastic case \cite{Mihai:2018a:MDWG,Mihai:2018b:MDWG,Mihai:2019a:MWG,Mihai:2019:MDWG}. For numerical methods applied to problems that are intractable analytically, we refer to \cite{Staber:2018:SG,Staber:2019:SGSMI}. Important applications include soft biological tissues (e.g., plants, arterial walls) and engineered structures (e.g., soft actuators) at large strains, where mathematical models that take into account the variability in the material responses are crucial. Similar modelling approaches can be developed for other mechanical systems.

\appendix
\section{Random shear moduli of stochastic anisotropic models}\label{sec:append}
In this appendix, we first state an important result concerning the summation of independent Gamma-distributed variables (see also Theorem~1 of \cite{Moschopoulos:1985}, where a proof is provided). This result is applicable  to linear combinations of independent Gamma-distributed random variables by rescaling. We  then apply this to derive the random shear moduli of the stochastic anisotropic model \eqref{eq:W:stoch} under small strains, as limiting cases of nonlinear shear moduli under large strains. The nonlinear elastic moduli presented here can be regarded as the anisotropic analogue of those discussed in \cite{Mihai:2017:MG} and \cite{Mihai:2018:MWG} for isotropic materials with deterministic and stochastic parameters, respectively. In the deterministic case, where the model parameters are single-valued constants, the linear shear moduli derived here are equivalent to those obtained through a different approach in \cite{Murphy:2014}.

\begin{theorem}\label{th:gammasum}.
If $\{R_{i}\}_{i=1,\cdots,n}$ are mutually independent Gamma-distributed random variables, with the corresponding shape and scale hyperparameters, $\rho_{1}^{(i)}$ and $\rho_{2}^{(i)}$, $i=1,2,\cdots, n$, respectively, such that $\rho_{2}^{(1)}=\min_{i=1,\cdots,n}\rho_{2}^{(i)}$, then the density of $R=\sum_{i=1}^{n}R_{i}$ can be expressed as follows,
	\begin{equation}
	g(R)=C\sum_{k=0}^{\infty}\frac{\delta_{k}R^{\rho+k-1}e^{-R/\beta_{1}}}{\beta_{1}^{\rho+k}\Gamma(\rho+k)}, \qquad R>0,
	\end{equation}
	where
	\begin{eqnarray}
	\rho&=&\sum_{i=1}^{n}\rho_{1}^{(i)},\\
	C&=&\prod_{i=1}^{n}\left(\frac{\rho_{2}^{(1)}}{\rho_{2}^{(i)}}\right)^{\rho_{1}^{(i)}},
	\end{eqnarray}
	and $\delta_{k}$ satisfies
	\begin{equation}
	e^{\sum_{k=1}^{\infty}\gamma_{k}\left(1-\rho_{1}^{(1)}t\right)^{-k}}=\sum_{k=0}^{\infty}\delta_{k}\left(1-\rho_{1}^{(1)}t\right)^{-k},
	\end{equation}
	with
	\begin{equation}
	\gamma_{k}=\frac{1}{k}\sum_{i=1}^{n}\rho_{1}^{(i)}\left(1-\frac{\rho_{2}^{(1)}}{\rho_{2}^{(i)}}\right)^{k}, \qquad k=1,2,\cdots.
	\end{equation}
	
\end{theorem}

\begin{figure}[htbp]
	\begin{center}
		\includegraphics[width=0.9\textwidth]{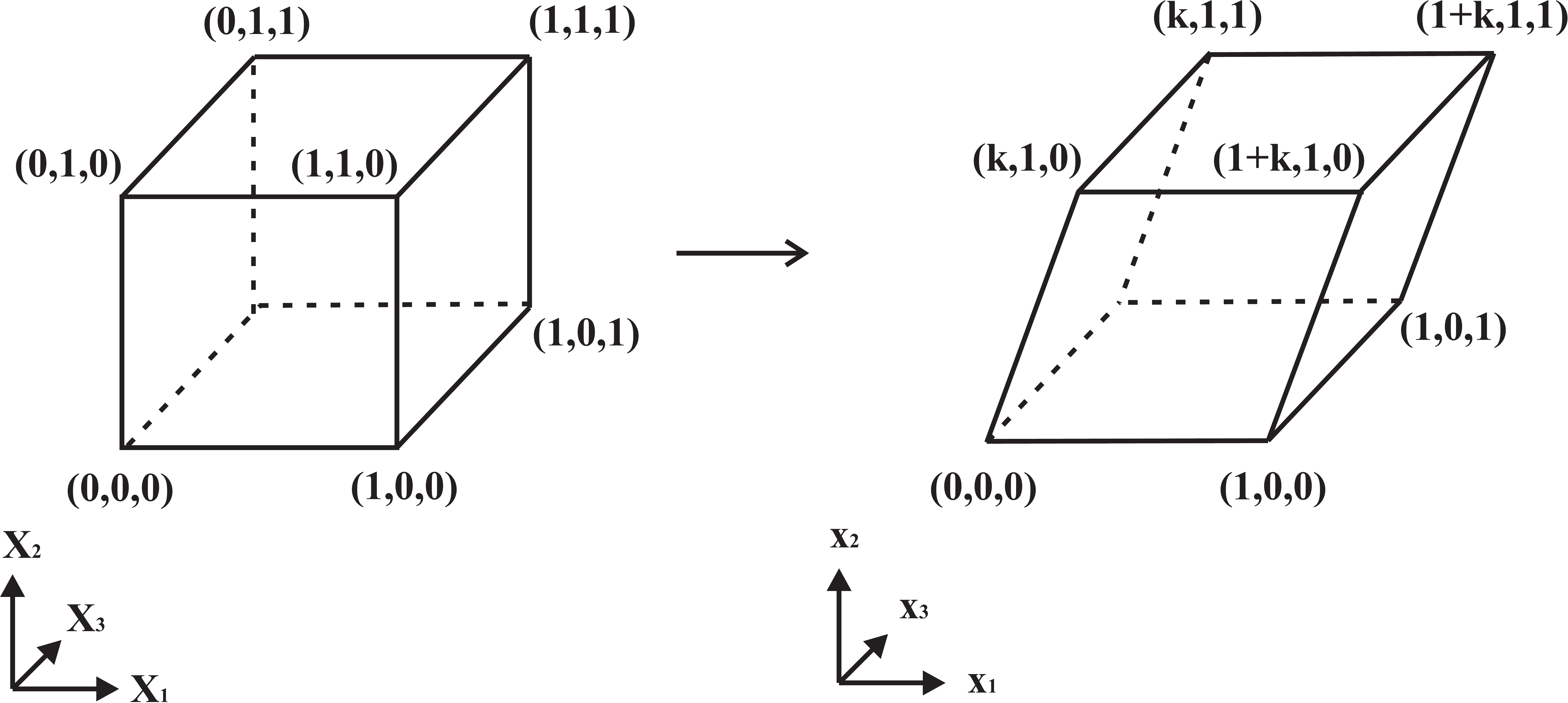}
		\caption{Cuboid (left) deformed by simple shear (right).}\label{fig:ss}
	\end{center}
\end{figure}

We now consider a cuboid sample of stochastic material, described by \eqref{eq:W:stoch}, and assume that the edges of the cuboid are aligned with the Cartesian axes in the reference configuration, while the two preferred directions are given by
\begin{equation}\label{eq:ss:M12}
\textbf{M}_{1}
=\left[
\begin{array}{c}
M_{11}\\
M_{12}\\
M_{13}
\end{array}
\right]
=\left[
\begin{array}{c}
\cos\Phi\\
\sin\Phi\\
0
\end{array}
\right]
\qquad\mbox{and}\qquad
\textbf{M}_{2}
=\left[
\begin{array}{c}
M_{21}\\
M_{22}\\
M_{23}
\end{array}
\right]
=\left[
\begin{array}{c}
-\cos\Psi\\
\sin\Psi\\
0
\end{array}
\right].
\end{equation}
First, we deform the cuboid by the simple shear
\begin{equation}\label{eq:ss:12}
x_{1}=X_{1}+kX_{2},\qquad x_{2}=X_{2},\qquad x_{3}=X_{3},
\end{equation}
where $(X_{1},X_{2},X_{3})$ and $(x_{1},x_{2},x_{3})$ are the  Cartesian coordinates for the reference and current configurations, respectively (see Figure~\ref{fig:ss}). For the deformation \eqref{eq:ss:12}, the gradient tensor is equal to
\begin{equation}\label{eq:ss:F}
\textbf{F}=\left[
\begin{array}{ccc}
1 & k & 0\\
0      & 1 & 0\\
0      & 0 & 1\\
\end{array}
\right],
\end{equation}
where $k>0$ is the shear parameter, representing the shear strain. The corresponding left and right Cauchy-Green deformation tensors are, respectively,
\begin{equation}\label{eq:ss:B}
\textbf{B}=\left[
\begin{array}{ccc}
1+k^{2} & k & 0\\
k      & 1 & 0\\
0      & 0 & 1\\
\end{array}
\right],\qquad
\textbf{C}=\left[
\begin{array}{ccc}
1 & k & 0\\
k      & 1+k^{2} & 0\\
0      & 0 & 1\\
\end{array}
\right].
\end{equation}
The two preferred directions, defined by \eqref{eq:ss:M12}, deform into the following directions, 
\begin{equation}\label{eq:ss:M12:st}
\textbf{m}_{1}=\textbf{F}\textbf{M}_{1}
=\left[
\begin{array}{c}
\cos\Phi+k\sin\Phi\\
\sin\Phi\\
0
\end{array}
\right],\qquad
\textbf{m}_{2}=\textbf{F}\textbf{M}_{2}
=\left[
\begin{array}{c}
-\cos\Psi+k\sin\Psi\\
\sin\Psi\\
0
\end{array}
\right].
\end{equation}
Then, the non-zero components of the associated stress tensor, given by \eqref{eq:stress}, with $\textbf{B}$ defined by \eqref{eq:ss:B}, take the form,
\begin{equation}\label{eq:ss:stress:nonzero}
\begin{split}
T_{11}&=-p+\beta_{1}\left(1+k^2\right)+\beta_{4}\left(\cos\Phi+k\sin\Phi\right)^2+\beta_{6}\left(\cos\Psi-k\sin\Psi\right)^2,\\
T_{12}&=\beta_{1}k+\beta_{4}\sin\Phi\left(\cos\Phi+k\sin\Phi\right)-\beta_{6}\sin\Psi\left(\cos\Psi-k\sin\Psi\right),\\
T_{22}&=-p+\beta_{1}+\beta_{4}\sin^2\Phi+\beta_{6}\sin^2\Psi,\\
T_{33}&=-p+\beta_{1},
\end{split}
\end{equation}
where
\begin{equation}\label{eq:ss:betas}
\begin{split}
\beta_{1}&=2\frac{\partial\mathcal{W}}{\partial I_{1}}=\mu,\\
\beta_{4}&=2\frac{\partial\mathcal{W}}{\partial I_{4}}=\mu_{4}\left(I_{4}-1\right)=\mu_{4}k\sin\Phi\left(k\sin\Phi+2\cos\Phi\right),\\
\beta_{6}&=2\frac{\partial\mathcal{W}}{\partial I_{6}}=\mu_{4}\left(I_{6}-1\right)=\mu_{6}k\sin\Psi\left(k\sin\Psi-2\cos\Psi\right).
\end{split}
\end{equation}
In this case, we can define the following nonlinear shear modulus \cite{Mihai:2017:MG}, as the ratio of the shear stress to the shear strain, under large strain,
\begin{equation}\label{eq:nlshearmod:T12}
\widetilde{\mu}_{12}(k)=\frac{T_{12}}{k}.
\end{equation}
By \eqref{eq:ss:stress:nonzero} and \eqref{eq:ss:betas}, the shear modulus given by \eqref{eq:nlshearmod:T12} takes the form
\begin{equation}\label{eq:nlshearmod:12}
\begin{split}
\widetilde{\mu}_{12}(k)&=\mu+\mu_{4}\sin^2\Phi\left(2\cos^2\Phi+3k\sin\Phi\cos\Phi+k^2\sin^2\Phi\right)\\
&+\mu_{6}\sin^2\Psi\left(2\cos^2\Psi-3k\sin\Psi\cos\Psi+k^2\sin^2\Psi\right).
\end{split}
\end{equation}
Under small shear, such that $k\to 0$, the modulus \eqref{eq:nlshearmod:12} converges to the linear elastic limit
\begin{equation}\label{eq:shearmod:12}
\mu_{12}=\lim_{k\to0}\widetilde{\mu}_{12}(k)=\mu+2\mu_{4}\sin^2\Phi\cos^2\Phi+2\mu_{6}\sin^2\Psi\cos^2\Psi.
\end{equation}
Next, we assume that the undeformed cuboid, with the preferred directions given by \eqref{eq:ss:M12}, is subjected to the simple shear deformation
\begin{equation}\label{eq:ss:13}
x_{1}=X_{1}+kX_{3},\qquad x_{2}=X_{2},\qquad x_{3}=X_{3}.
\end{equation}
Following analogous calculations to those detailed above, we can define the nonlinear shear modulus
\begin{equation}\label{eq:nlshearmod:T13}
\widetilde{\mu}_{13}(k)=\frac{T_{13}}{k},
\end{equation}
and obtain
\begin{equation}\label{eq:nlshearmod:13}
\widetilde{\mu}_{13}(k)=\mu.
\end{equation}
As $\mu$ is constant, the shear modulus described by \eqref{eq:nlshearmod:13} coincides with its linear elastic limit
\begin{equation}\label{eq:shearmod:13}
\mu_{13}=\lim_{k\to0}\widetilde{\mu}_{13}(k)=\mu.
\end{equation}
Finally, assuming that the original cuboid deforms by the simple shear
\begin{equation}\label{eq:ss:23}
x_{1}=X_{1},\qquad x_{2}=X_{2}+kX_{3},\qquad x_{3}=X_{3},
\end{equation}
while the preferred directions are given by \eqref{eq:ss:M12}, we define the nonlinear shear modulus
\begin{equation}\label{eq:nlshearmod:T23}
\widetilde{\mu}_{23}(k)=\frac{T_{23}}{k}.
\end{equation}
In this case also, we obtain
\begin{equation}\label{eq:nlshearmod:23}
\widetilde{\mu}_{23}(k)=\mu,
\end{equation}
with its linear elastic limit
\begin{equation}\label{eq:shearmod:23}
\mu_{23}=\lim_{k\to0}\widetilde{\mu}_{23}(k)=\mu.
\end{equation}
For the deterministic problem, the formulae of the shear moduli \eqref{eq:shearmod:12}, \eqref{eq:shearmod:13}, \eqref{eq:shearmod:23}, under infinitesimal deformation, are equivalent to those described by equation (6.1) of \cite{Murphy:2014}. In addition, for the stochastic problem, the material parameters are random variables defined by probability distribution. In particular, if $\mu$, $\mu_{4}$, and $\mu_{6}$ are statistically independent, and each of them is drawn from a Gamma distribution, then, by Theorem~\ref{th:gammasum}, one can obtain the probability distributions of the shear moduli described by \eqref{eq:shearmod:12}, \eqref{eq:shearmod:13} and \eqref{eq:shearmod:23}, respectively.

\paragraph{Acknowledgement.} The support by the Engineering and Physical Sciences Research Council of Great Britain under research grants EP/R020205/1 for Alain Goriely and EP/S028870/1 for L. Angela Mihai is gratefully acknowledged.


\end{document}